\documentclass[10pt,aps,twocolumn,showpacs,superscriptaddress,floatfix,nofootinbib]{revtex4-2}
\pdfoutput=1
\usepackage[utf8]{inputenc}
\usepackage[english]{babel}
\usepackage[T1]{fontenc}
\usepackage{csquotes}
\usepackage{physics}
\usepackage{amsmath,amsfonts}
\usepackage[dvipsnames]{xcolor}
\usepackage[colorlinks=true]{hyperref}
\usepackage[capitalise]{cleveref}
\usepackage{listings}
\usepackage{graphicx}
\usepackage{blkarray} 
\usepackage{parskip}
\usepackage{multirow}
\graphicspath{{figures/}}
\usepackage{bm}

\begin{document}

\title{Phase diagram and crystal melting of helium-4 in two dimensions}

\author{David Linteau}
\email{david.linteau@epfl.ch}
\affiliation{Institute of Physics, \'{E}cole Polytechnique F\'{e}d\'{e}rale de Lausanne (EPFL), CH-1015 Lausanne, Switzerland}
\affiliation{Center for Quantum Science and Engineering, \'{E}cole Polytechnique F\'{e}d\'{e}rale de Lausanne (EPFL), CH-1015 Lausanne, Switzerland}

\author{Gabriel Pescia}
\affiliation{Institute of Physics, \'{E}cole Polytechnique F\'{e}d\'{e}rale de Lausanne (EPFL), CH-1015 Lausanne, Switzerland}
\affiliation{Center for Quantum Science and Engineering, \'{E}cole Polytechnique F\'{e}d\'{e}rale de Lausanne (EPFL), CH-1015 Lausanne, Switzerland}

\author{Jannes Nys}
\affiliation{Institute of Physics, \'{E}cole Polytechnique F\'{e}d\'{e}rale de Lausanne (EPFL), CH-1015 Lausanne, Switzerland}
\affiliation{Center for Quantum Science and Engineering, \'{E}cole Polytechnique F\'{e}d\'{e}rale de Lausanne (EPFL), CH-1015 Lausanne, Switzerland}

\author{Giuseppe Carleo}
\affiliation{Institute of Physics, \'{E}cole Polytechnique F\'{e}d\'{e}rale de Lausanne (EPFL), CH-1015 Lausanne, Switzerland}
\affiliation{Center for Quantum Science and Engineering, \'{E}cole Polytechnique F\'{e}d\'{e}rale de Lausanne (EPFL), CH-1015 Lausanne, Switzerland}

\author{Markus Holzmann}
\affiliation{Univ. Grenoble Alpes, CNRS, LPMMC, 38000 Grenoble, France}

\begin{abstract}

We study the zero-temperature phase diagram of two-dimensional helium-4 using neural quantum states.
Our variational description allows us to address liquid and solid phases using the same functional form as well as exploring possible melting scenarios, for instance via an intermediate hexatic phase. 
Notably, this is achieved by performing fixed pressure variational Monte Carlo calculations.
Within the isobaric ensemble framework, we are able to clearly identify the first-order liquid-solid phase transition.
However, in an intermediate region of nearly constant pressure, we find that simulations of $N=30$ atoms continuously transition from liquid to solid, with signatures of a hexatic order coexisting with a small condensate fraction.
Calculations for larger systems follow the metastable liquid and solid branches in this transient region.
We additionally compute the R\'enyi-2 entanglement entropy across the liquid-solid phase transition and find a sharp decrease upon freezing.

\end{abstract}

\maketitle

\paragraph*{Introduction--}
Unlike most materials, helium-4 does not freeze at ambient pressure, even when cooled down to absolute zero-temperature, but instead transitions to a superfluid state.
The discovery of superfluidity in liquid helium-4 \cite{1938_kapitza_superfluidity_helium4, 1938_allen_misener_superfluidity_helium4} motivated the description of this dissipationless fluid behavior \cite{1938_london_superfluidity_as_BEC, 1941_landau_superfluid_helium, 1947_bogolyubov_superfluidity, 1954_feynman_liquid_helium}, making helium-4 an emblematic quantum fluid. 
In turn, at sufficiently large pressure, helium-4 will transition to an ordinary solid phase, thus opening the possibility for an intermediate supersolid state, exhibiting superfluid properties while maintaining a crystalline order \cite{1969_andreev_lifshitz_supersolid_postulate,1969_thouless_flow_of_a_dense_superfluid,1970_leggett_can_a_solid_be_superfluid,1970_chester_presence_supersolidity,2012_boninsegni_supersolid_review}.
Although initial experimental evidence \cite{2004_kim_chan_supersolid_evidence_1,2004_kim_chan_supersolid_evidence_2,2006_kim_chan_supersolid_evidence_3} has been discarded theoretically \cite{2004_ceperley_bernu_no_supersolid,2006_clark_odlro_he4,2006_boninsegni_superglass_he4,2006_boninsegni_vacancy_induced_supersolidity_he4} and also experimentally \cite{2012_kim_chan_absence_of_supersolidity},
helium-4 still manifests unexplained properties, like anomalous plasticity in the bulk \cite{2020_beamish_review_plasticity}. 
Supersolid cluster phases have been predicted in different systems \cite{2010_cinti_supersolid_droplet_crystal,2011_saccani_supersolidity_soft_core_bosons,2012_saccani_excitation_spectrum_supersolid}
and apparent supersolidity has been observed in confined quantum gases \cite{2019_bottcher_supersolidity_qm_gas,2019_chomaz_supersolidity_qm_gas,2019_tanzi_supersolidity_qm_gas}, re-opening the debate concerning a supersolid phase in two-dimensional helium-4 \cite{2016_nakamura_liquid_crystal_he,2021_choi_modulated_superfluid_he4}.

Following the groundbreaking work of Kosterlitz and Thouless on thermal melting of solids in two dimensions \cite{1973_kosterlitz_thouless_melting_solids_2d}, Halperin, Nelson and Young \cite{1978_halperin_nelson_hexatic_1, 1979_halperin_nelson_hexatic_2, 1979_young_melting_coulomb_gas_2d} theorized the presence of an intermediate state between the liquid and solid phases conserving orientational order without translational order, the so-called hexatic phase.
While the hexatic phase has been observed in classical simulations \cite{2011_bernard_hexatic,2015_kapfer_2d_melting,2019_hajibabaei_2d_melting,2017_anderson_2d_melting}, its role in the melting of quantum crystals is largely unexplored.
Zero-temperature attempts using diffusion Monte Carlo (DMC) methods \cite{2008_apaja_hexatic_he4} have been biased so far by the trial wavefunction functional form explicitly imposing a hexatic or crystal symmetry.

In this work, we explore the phase diagram of helium-4 confined to two spatial dimensions at zero-temperature, using a neural quantum state (NQS) trial wavefunction \cite{2017_carleo_nqs} combined with a deep layer structure in continuous space \cite{2015_taddei_iterative_backflow_fermi_liquids,2019_holzmann_backflow,2020_hermann_neural_net_electrons, 2020_pfau_deep_net}. 
Using the same symmetric and translationally invariant functional form, we are able to accurately describe phases of different spatial symmetries (liquid and solid), as well as the phenomena of Bose-Einstein condensation. 
In contrast to previous numerical studies based on variational (VMC) and diffusion Monte Carlo \cite{1998_gordillo_phase_diagram_he4, 1998_moroni_he4_excitation_spectrum, 2000_krishnamachari_shadow_wavefunction, 2013_arrigoni_excitation_spectrum_2d}, 
we perform fixed-pressure variational Monte Carlo calculations. 
The accuracy of our description allows us to properly address the (putative first-order) liquid-solid phase transition as well as possible intermediate exotic phases \cite{2008_apaja_hexatic_he4,2020_gordillo_superfluid_supersolid_he4,2020_boninsegni_specific_heat_he4_graphite}. 
While our simulations of small cells reveal some sign of an intermediate solid or hexatic phase with an anomalously large condensate fraction, simulations of larger systems lack signatures of these phases, indicating strong finite-size bias
and questioning the appearance of these exotic phases in the thermodynamic limit.
We further study the entanglement entropy of the system, showing that it grows with increasing pressure before dropping sharply in the transition region.

\begin{figure*}[ht]
    \centering
    \centering
    \includegraphics[width=\textwidth]{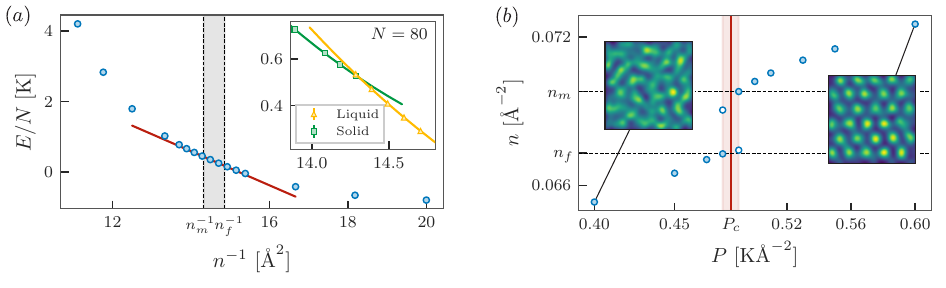}
  \caption{
  Phase diagram showcasing the first-order phase transition of helium-4 in two dimensions at absolute zero shown in the canonical ensemble (a) and the isobaric ensemble (b).
  In both ensembles, $N=30$ particles were simulated.
  Dashed lines were added at $n_f \approx 0.0673$ \AA$^{-2}$ and $n_m \approx 0.0698$ \AA$^{-2}$ to roughly identify the position of the freezing and melting densities respectively.
  The critical pressure is also roughly approximated to $P_c \approx 0.485(5)$ K\AA$^{-2}$, with the red line being the associated isobaric line.
  The uncertainty on $P_c$ is taken to be that of the chosen pressure grid size.
  (a) The approximate liquid-solid coexistence region is shaded in grey.
  For comparison, the $N=80$ data is shown in the inset with two distinct liquid and solid metastable branches forming.
  (b) The insets display the expectation value of the density operator both in the liquid and solid phases.
  Metastable states are shown as hollow points.
  }
  \label{fig:phase_diagrams}
\end{figure*}

\paragraph*{Method--}
The Hamiltonian for $N$ helium-4 atoms of mass $m$ is given by
\begin{equation} \label{eq:hamiltonian}
    H = - \frac{\hbar^2}{2m} \sum_{i=1}^N \nabla_{\mathbf{r}_i}^2 + \sum_{i<j} v(|\mathbf{r}_i-\mathbf{r}_j|),
\end{equation}
where we choose the HFDHE2 pair potential for $v(r)$ \cite{1979_aziz_potential}.
Periodic boundary conditions are used to describe bulk helium at a number density $n=N/V$, where $V$ is the volume of the simulation cell.

We construct the bosonic trial wavefunction for the ground state of helium-4 using two building blocks: (1) a short-range part of the McMillan form \cite{1965_mcmillan_ground_state_he4}, and (2) a ``long-range'' part which accounts for the rest of the correlation, modeled by a MP-NQS~\cite{2024_pescia_mpnn_electrongas}. 
Putting these two pieces together, the resulting wavefunction is
\begin{equation} \label{eq:trial wavefunction}
    \Psi_{\boldsymbol{\theta}}(\mathbf{R}) = \prod_{i<j} \exp[\theta_1 \phi(\mathbf{e}_{ij}^{(K)})] \exp \left[ - \frac{\theta_2}{d_\text{sin}(\mathbf{r}_i, \mathbf{r}_j)^{5}} \right],
\end{equation}
where $\mathbf{R} = (\mathbf{r}_1, \hdots, \mathbf{r}_N) \in \mathbb{R}^{N \cross 2}$ denotes all $N$ single-particle coordinates and the periodized distance $d_\text{sin}(\mathbf{r}_i, \mathbf{r}_j) = |(\mathbf{L}/2) \sin (\pi \mathbf{r}_{ij}/ \mathbf{L})|$ is introduced, with $\mathbf{L}=(L_x,L_y)$~\cite{2022_pescia_periodic_continuous_systems}.
The GNN implementation maps the single-particle coordinates to backflowed coordinates $\mathbf{e}_{ij}^{(K)}$, that is $\mathrm{GNN}: \mathbf{R} \to \{\mathbf{e}_{ij}^{(K)}\}$, where the superscript $K$ specifies the number of message-passing iterations performed within the GNN.
The function $\phi$ is parameterized with a multi-layer perceptron.
The weights of $\phi$ and $\mathbf{e}_{ij}^{(K)}$, along with $\theta_1$ and $\theta_2$, are variational parameters encapsulated in the vector $\boldsymbol{\theta}$.
By construction, our final trial wavefunction is invariant under the permutation of single-particle coordinates and respects periodic boundary conditions (see Supplemental Material for further implementation details and performance).

Our ansatz is general, and can be justified by the mean value theorem for integrals \cite{2024_pescia_mpnn_electrongas}, and similarly to the iterated backflow construction \cite{2018_ruggeri_backflow}, convergence to the
exact ground state can be reached. 

\paragraph*{Results--}
We perform simulations at zero-temperature in the canonical and isobaric ensembles. 
In the canonical ensemble, $N$ and $V$ are fixed.
In that case, we use a rectangular simulation cell compatible with the expected triangular solid \cite{1998_gordillo_phase_diagram_he4}, and minimize the expectation value of the Hamiltonian,  given in \cref{eq:hamiltonian}, to obtain the system's internal energy as a function of density, denoted $E(n)$.
In the isobaric ensemble, $N$ and $P$ are fixed.
In this case we minimize the Gibbs free energy (or enthalpy since $T = 0$) given by
\begin{equation}
    G = H + P V,
\end{equation}
where $V = L_x L_y$ (in 2D).
The lengths $L_x$ and $L_y$ that characterize the simulation cell can then be treated as variational parameters, similarly to those that appear in the trial wavefunction, as detailed in the End Matter.

In \cref{fig:phase_diagrams} (a), we show the equation of state obtained from our calculations in the canonical ensemble with $N=30$.
Notably, the constant slope of the energy indicates a zone of constant pressure between $n_f$ and $n_m$ (hence the isobaric red line), which are respectively the freezing density of the liquid and the melting density of the solid.
This behavior is expected for a first-order phase transition, with the interval $[n_f,n_m]$ defining the liquid-solid co-existence region (i.e., the shaded gray region).
In contrast to previous calculations \cite{1998_gordillo_phase_diagram_he4, 2000_krishnamachari_shadow_wavefunction, 2013_arrigoni_excitation_spectrum_2d}, our variational wavefunction does not follow the metastable branches at $N=30$. 
Instead, metastability only shows up in our larger simulations with $N=80$, as displayed in the inset.

\begin{figure}[t]
    \centering
    \includegraphics[width=\columnwidth]{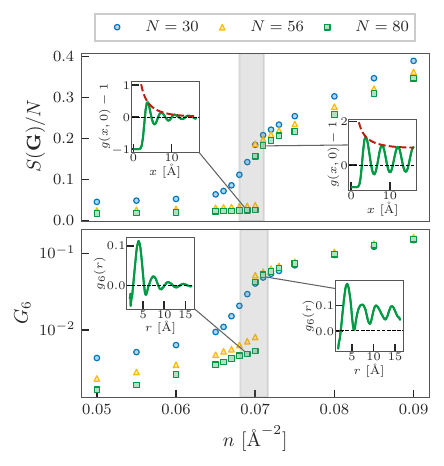}
  \caption{
  Translational and orientational orders across the liquid-solid phase transition to probe a possible hexatic phase.
  In the top panel, the structure factor evaluated at the reciprocal lattice vector $\mathbf{G}=4\pi/(a\sqrt{3})(0,1)$ is shown, with $a$ the lattice constant.
  Pair correlation function insets in the liquid and solid phases are also shown.
  In the bottom panel, the hexatic correlation function $g_6(r)$ and the integrated hexatic correlation function $G_6$ are shown.
  In the top panel, a $1/x$ (periodized) fit, depicted as a red dashed line, is performed on the peaks of the pair correlation function $g(x,0)$, where $x$ is the spatial coordinates in the $x$-direction. 
  The fit is explicitly given by $f(x)=c_1+c_2(1/x+1/(\text{min}(\mathbf{L})-x))$ for some fitting parameters $c_1$ and $c_2$. 
  In the solid, the peaks in $g(x,0)$ saturate at a constant $c_1 \ne 0$ (up to a $1/x$ correction), while in the liquid $c_1 \to 0$.
  For large systems, $g_6(r)$ either decays exponentially to 0 (liquid) or plateaus at a non-zero value (solid).
  In particular, it does not decay algebraically, which would suggest a hexatic phase.
  The liquid-solid coexistence region, shaded in grey, was obtained using the Maxwell construction for $N=80$ (see the Supplemental Material for more details).
  }
  \label{fig:translational_and_orientational_order}
\end{figure}

Simulations of solid structures are sensitive to the shape of the simulation cell, potentially biasing calculations in the canonical ensemble.
In order to enable structural relaxation, we have performed fixed pressure calculations shown in \cref{fig:phase_diagrams} (b).
The abrupt change in density supports the first-order character of the transition, with the resulting ratio $L_y/L_x$ compatible with the triangular lattice forming at high pressures (see the Supplemental Material for simulations with variable shapes using non-rectangular cells).
The associated critical pressure $P_c \approx 0.485(5)$ K\AA$^{-2}$ agrees within our resolution with the one of the canonical ensemble simulations, and is also close to the Path Integral Monte Carlo results of \cite{1998_gordillo_phase_diagram_he4} at low temperatures.

So far, our results indicate a first-order phase transition that occurs at a critical pressure $P_c$, from the low density liquid phase $n \le n_f$ to the solid phase at high density $n \ge n_m$.
The question of how two-dimensional classical solids melt as temperature increases -- whether through a first-order transition or continuously through an intermediate hexatic phase -- remains a fundamental problem that has been challenging classical simulations for several decades \cite{2011_bernard_hexatic,2015_kapfer_2d_melting,2019_hajibabaei_2d_melting,2017_anderson_2d_melting}.
In order to analyze if or how such scenarios takes place in the quantum realm, and in particular at zero-temperature, we turn back to fixed density simulations, performed for different system sizes, namely $N=30, 56$ and $80$.

Information about structural properties is encoded in the pair-correlation function $g(\mathbf{r})$.
The crystalline order is characterized by Bragg peaks in its Fourier transform, the structure factor $S(\mathbf{k})$. 
The pronounced Bragg peak at the reciprocal lattice vector $\mathbf{G}=4 \pi / (a \sqrt{3})(0,1)$ of the hexagonal lattice, shown in Fig.~\ref{fig:translational_and_orientational_order}, supports the onset of crystalline order close to $n_m$, also directly visible in the pair correlation function showing long-range positional order. The slow $1/x$ decay of the maxima approaching the non-vanishing asymptotic value in the solid phase is compatible with the behavior of the structure factor close to the Bragg peaks, $S(\mathbf{G}+\mathbf{q}) \sim |\mathbf{q}|^{-1}$ for $q \to 0$, and entails a $1/N^{1/2}$ finite size corrections for the amplitude of the Bragg peak.  
Whereas for $N=30$ the positional order vanishes smoothly between $n_m$ and $n_f$, the metastability of larger systems leads to an abrupt disappearance of positional order.

\begin{figure}[!t]
    \centering
    \includegraphics[width=\columnwidth]{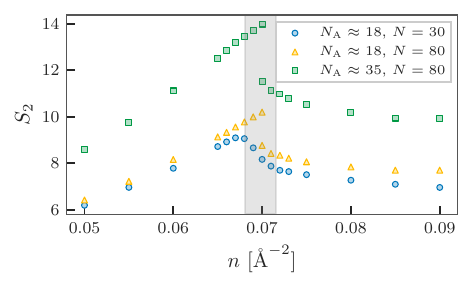}
    \caption{
    R\'enyi-2 entanglement entropy using a spherical partition with a fixed number of particles $N_\text{A}$ inside.
    }
    \label{fig:s2_vs_n}
\end{figure}

To probe the presence of a possible hexatic phase in the intermediate region, we further calculate the hexatic correlation function 
\begin{equation} \label{eq:g6}
    g_6(r) = \mathbb{E}[\psi_6^*(\mathbf{r}_i) \psi_6(\mathbf{r}_j)],
\end{equation}
where $\psi_6(\mathbf{r}_i)$ is the hexatic orientational order parameter defined as 
\begin{equation} \label{eq:psi6}
    \psi_6(\mathbf{r}_i) = \frac{1}{6} \sum_{j \in \mathcal{N}_6(i)} e^{6 i \theta_{ij}},
\end{equation}
and the average in \cref{eq:g6} is taken over particle-pairs $(i,j)$ such that $r = |\mathbf{r}_{ij}|$.
The sum in \cref{eq:psi6} is taken over the six nearest neighbors of particle $i$, and $\theta_{ij}$ is the angle between $\mathbf{r}_{ij} = \mathbf{r}_i - \mathbf{r}_j$ and a fixed arbitrary vector, say $(1,0)$.
\begin{figure}[!t]
    \centering
    \includegraphics[width=\columnwidth]{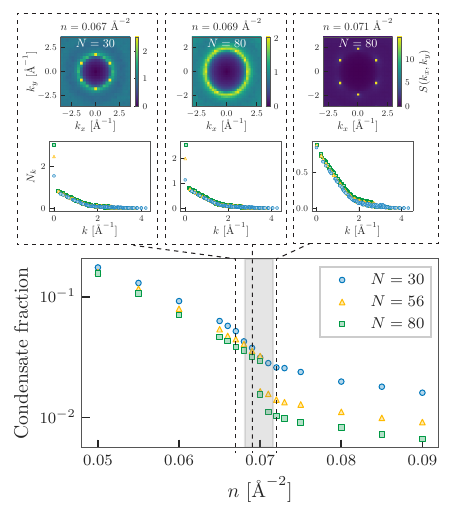}
    \caption{
    Investigating off-diagonal long-range order and supersolidity with the condensate fraction $N_0/N$.
    The insets show the momentum distribution $N_k$ (as a function of $k \equiv \sqrt{k_x^2 + k_y^2}$) and the structure factor $S(k_x,k_y)$ in the vicinity of the liquid-solid coexistence region, at $n=0.067$ \AA$^{-2}$ with $N=30$ atoms, and $n=0.069, 0.071$ \AA$^{-2}$ for $N=80$ atoms.
    }
    \label{fig:condensate_fraction_vs_n}
\end{figure}
From the integrated value of the hexatic correlation function, $G_6 = (2\pi/V) \int_0^{\mathrm{min}(\mathbf{L})/2} r dr g_6(r)$, shown in \cref{fig:translational_and_orientational_order}, we see that orientational order seems to follow positional order with a similar sudden suppression for larger systems.
Our data therefore support a first-order melting transition at zero-temperature.

The entanglement content of a system can provide additional insight into the phase transition. 
Here, we focus on the bipartite entanglement between two spatial subregions of the simulation cell, referred to as partition A and partition B, and explicitly defined below.
The reduced density matrix for partition A is obtained by tracing out the degrees of freedom of partition B, that is, $\rho_\text{A} = \text{Tr}_\text{B}\rho$, where $\rho = |\Psi\rangle \langle \Psi|/\langle \Psi | \Psi \rangle$ and $|\Psi\rangle$ is the ground state wavefunction.
We compute the so-called R\'enyi-2 entropy
\begin{equation} \label{eq:renyi-2_definition}
    S_2 = -\ln(\text{Tr}_\text{A} \rho_\text{A}^2),
\end{equation}
based on the Monte Carlo estimator previously discussed in \cite{2010_hastings_renyi2, 2012_tubman_renyi2_vmc, 2013_mcminis_renyi2_fermions, 2014_herdman_renyi2_pimc, 2023_sinibaldi_unbiasing, 2024_mauron_topological_entanglement} (see the End Matter section).
In \cref{fig:s2_vs_n}, we show $S_2$ for a spherical bipartitioning of the simulation cell, where partition A is a surface  of area $\pi R^2$ containing $N_\text{A} \approx 18$ or $35$ atoms. 
The entanglement increases with density in the liquid phase and then sharply drops at the onset of the presumed first-order phase transition.

At low temperature, liquid helium-4 turns superfluid via a Kosterlitz-Thouless transition \cite{1978_bishop_superfluid_2d_he4} to an algebraically ordered phase \cite{1972_berezinskii_destruction_of_lro,1973_kosterlitz_thouless_melting_solids_2d,1974_kosterlitz_2d_xy_model}. At zero-temperature, true off-diagonal long-range order (ODLRO) is expected in the liquid phase. 
In the presence of ODLRO, the one-particle density matrix $n(\mathbf{r}, \mathbf{r}')=\langle \Psi_0 | \hat{\psi}^\dagger(\mathbf{r}) \hat{\psi}(\mathbf{r}') |\Psi_0 \rangle$
of the ground state $\Psi_0$, where $\hat{\psi}(\mathbf{r})$ ($\hat{\psi}^\dagger(\mathbf{r})$) is the annihilation (creation) operator of a particle at $\mathbf{r}$, reaches a plateau for large separations, i.e. $\lim_{|\mathbf{r}-\mathbf{r}'| \to \infty} n(\mathbf{r}, \mathbf{r}') = n_0$ where $n_0$ is the condensate density.
Using translational invariance,
the associated Monte Carlo estimator can be written as a ratio of the ground state wavefunction \cite{1965_mcmillan_ground_state_he4} evaluated at the same spatial configuration except for one particle displaced by a vector $\mathbf{r}$, that is 
\begin{equation}
    \frac{n(\mathbf{r})}{n} =  \mathbb{E}\left[ \frac{\Psi_0(\mathbf{r}_1 + \mathbf{r}, \mathbf{r}_2, \hdots, \mathbf{r}_N)}{ \Psi_0(\mathbf{r}_1, \mathbf{r}_2, \hdots, \mathbf{r}_N) } \right].
\end{equation}
A non-vanishing condensate density will lead to a peak in the Fourier transform, the momentum distribution, 
\begin{equation}
    N_\mathbf{k} = \int d \mathbf{r} e^{i \mathbf{k} \cdot \mathbf{r}} n(\mathbf{r}),
\end{equation}
with an extensive population in the $k=0$ mode which is not analytically connected to the others $k \ne 0$ modes featuring instead $N_{k \ne 0} \sim {\cal O}(1)$.

From \cref{fig:condensate_fraction_vs_n}, Bose condensation clearly emerges in the liquid ($n \le n_f$), whereas $N_0$ remains of order one on the solid side ($n \ge n_m$), which is compatible with the analytical continuation $N_{k \ne 0}  \to N_0$ for $k \to 0$.
We further note that, apart from $k=0$, the momentum distributions of the different system sizes seem to collapse on a common line that describes the thermodynamic limit values of $N_{k\ne 0}$. 
However, the non-analytic behavior $\sim k^{-1}$ of $N_{k \to 0}$ in the liquid triggers important finite-size effects of the condensate density $n_0 = N_0/V = n - V^{-1} \sum_{\mathbf{k} \ne 0} N_k$, which are clearly visible in \cref{fig:condensate_fraction_vs_n}. 
To leading order, finite size effects are given by $[ \int d^2k/(2 \pi)^2 - V^{-1} \sum_{k \ne 0} ] N_k \sim N^{-1/2}$ \cite{2016_holzmann_finite_size_effects}. 

Our data is therefore fully consistent with the absence of ODLRO in the solid phase with long-range crystalline order in the thermodynamic limit.
However, inside the (presumed) mixture region, between $n_f$ and $n_m$, for our $N=30$ system, apparent supersolid behavior can be observed. Within this region, $N_0$ remains visibly disconnected from $N_{k \to 0}$ while orientational and translational orders increase smoothly when crossing over from the liquid to the solid phase via an intermediate hexatic region.

\paragraph*{Conclusion}
We have studied the zero-temperature phase diagram of helium-4 using a highly flexible and accurate neural quantum state to represent the ground state wavefunction.
Our calculations at zero-temperature are consistent with a first-order phase transition from a superfluid to a normal (non-Bose condensed) quantum solid. 
Interestingly, the sharp decrease of the entanglement entropy observed at freezing indicates a zero-temperature analog of the entropy drop seen at a finite-temperature first-order transition.
However, for small systems, despite imposing periodic boundary conditions, we still observe a crossover to an intermediate solid or hexatic phase with an anomalously large condensate fraction. 
Therefore, one might expect such a crossover to occur in strongly inhomogeneous systems or nanodroplets \cite{2001_toennies_superfluid_he_droplets} around freezing densities. 
The isobaric ensemble methodological framework presented here, based on NQS, enables structural optimizations that can achieve diffusion Monte Carlo precision, which could be especially relevant for electronic structure calculations.

\paragraph*{Acknowledgments--}
The authors thank Saverio Moroni for discussions and comments on the manuscript.
G. C. acknowledges insightful discussions with Sergey Bravyi, who pointed out the improved overlap estimator devised by Marteen Van den Nest, adopted by us for the calculation of the Rényi entropy.
The simulations in this work were carried out using NetKet \cite{2019_netket,2022_netket_3}, which is based on Jax \cite{2018_jax} and MPI4Jax \cite{2021_hafner_mpi4jax}.
The authors acknowledge support from SEFRI under Grant No.\ MB22.00051 (NEQS - Neural Quantum Simulation) and from the French Agency for Research, project SIX (ANR-23-CE30-0022).

\section*{End Matter} 

In this section, we first outline our variational Monte Carlo implementation in the isobaric ensemble.
In the second part, we introduce an improved estimator for the calculation of the R\'enyi-2 entropy in continuous space -- which significantly reduces the variance -- and detail its implementation in the crystalline phase.

\paragraph*{Isobaric Ensemble--}
Let $\boldsymbol{\alpha}=\{L_x,L_y,\theta\}$ be the set of variational parameters characterizing the size and shape of the simulation cell. 
The trial wavefunction $\Psi_{\boldsymbol{\theta}}(\mathbf{R})$ depends in principle on all the variational parameters $\boldsymbol{\theta}$, with $\boldsymbol{\alpha}$ being a subset, that is $\boldsymbol{\alpha} \subset \boldsymbol{\theta}$.
However, we find that it is much more convenient in practice to pass all the geometrical dependency $\boldsymbol{\alpha}$ to the Gibbs operator, and remove all of it from the trial wavefunction.
This can be achieved by sampling unit scaled coordinates $\mathbf{S} = (\mathbf{s}_1, \hdots, \mathbf{s}_N) \in ([0,1] \times [0,1])^N$ instead of the physical coordinates $\mathbf{R}=(\mathbf{r}_1,\hdots,\mathbf{r}_N) \in ([0,L_x] \times [0,L_y])^N$. 
These physical coordinates can be obtained by the following affine transformation of the unit square
\begin{equation} \label{eq:affine_transformation_square_parallelogram}
    \begin{cases}
        x = L_x m + \cos \theta L_y n & \\
        y = \sin\theta L_y n &
    \end{cases}
    ,
\end{equation}
where $\mathbf{s} \equiv (m,n) \in [0,1]\times[0,1]$.
The affine transformation thus converts a (dimensionless unit) single-particle coordinate $\mathbf{s}_i$ to the associated physical single-particle coordinate $\mathbf{r}_i$, where $i$ is the particle index ranging from 1 to $N$.

As a result, the kinetic and potential energy parts of the local Gibbs energy estimator $\tilde{G}_\text{loc}(\mathbf{S})$ can be written in terms of the unit coordinates as follows 
\begin{align} \label{eq:k_and_v_gibbs_local_estimators}
    \tilde{K}_\text{loc}(\mathbf{S}) 
    &= -\frac{1}{2} \sum_{i=1}^N \bigg[ \sum_{\alpha,\beta=1}^d g^{\alpha \beta} 
    \frac{ \partial^2 }{\partial \mathbf{s}_i^\alpha \partial \mathbf{s}_i^\beta} \log \Psi(\mathbf{S}) \nonumber \\ 
    &\quad + ((J^{-1})^\text{T} \nabla_{\mathbf{s}_i} \log \Psi(\mathbf{S}) )^2 \bigg], \nonumber \\
    \tilde{V}_\text{loc}(\mathbf{S}) &= V(J \mathbf{S}) + P L_x L_y \sin \theta.
\end{align}
where $J \mathbf{S} \equiv (J \mathbf{s}_1, \hdots, J \mathbf{s}_N)$, with $J$ and $g^{\alpha \beta}$ being the Jacobian and the inverse metric tensor of the affine transformation in \cref{eq:affine_transformation_square_parallelogram}, given respectively by
\begin{align}
    J &= 
    \begin{pmatrix}
        L_x & \cos \theta L_y \\
        0 & \sin \theta L_y
    \end{pmatrix}
    , \\
    g^{\alpha \beta} &=
    \begin{pmatrix}
        \frac{ \text{csc}^2\theta }{ L_x^2} & -\frac{\text{cot}\theta \text{csc}\theta}{L_x L_y} \\
        -\frac{\text{cot}\theta \text{csc}\theta}{L_x L_y} & \frac{ \text{csc}^2\theta }{ L_y^2}
    \end{pmatrix}
    .
\end{align}
The expressions given in \cref{eq:k_and_v_gibbs_local_estimators} simplify the computer implementation since the Monte Carlo configurations $\mathbf{S}$ can be sampled from the unit box $([0,1] \times [0,1])^N$ independently of the varying geometrical parameters $L_x$, $L_y$ and $\theta$.

The dependence on the variational parameters $\boldsymbol{\alpha}$ in the Gibbs operator must be taken into account when taking gradients.
Given an arbitrary operator $A(\boldsymbol{\theta})$ that depends on the variational parameters $\boldsymbol{\theta}$, the general expression for the gradient entries reads 
\begin{equation} \label{eq:gradient_wrt_a_var_param}
    \frac{\partial \langle A(\boldsymbol{\theta}) \rangle}{\partial \theta_i} 
    = 2 \text{Re}[f_i] + \mathbb{E}[\partial_{\theta_i} A_\text{loc}(\boldsymbol{\theta})],
\end{equation}
where the usual force term, given by a covariance
\begin{equation}
    f_i = \text{Cov}[O_i, A_\text{loc}(\boldsymbol{\theta})] = \mathbb{E}[O_i^*(A_\text{loc}(\boldsymbol{\theta})-\mathbb{E}[A_\text{loc}(\boldsymbol{\theta})])],
\end{equation}
with $O_i(\mathbf{R}) = \partial_{\theta_i} \ln \Psi_{\boldsymbol{\theta}}(\mathbf{R})$, is supplemented by the term $\mathbb{E}[\partial_{\theta_i} A_\text{loc}(\boldsymbol{\theta})]$.

\paragraph*{Entanglement Entropy--} For the calculation of the Rényi-2 entropy, defined in \cref{eq:renyi-2_definition}, the so-called swap estimator, 
\begin{equation} \label{eq:swap_estimator}
    \text{Tr}_\text{A} \rho_\text{A}^2 
    = \mathbb{E}_{\substack{\mathbf{x}_1 \sim |\Psi(\mathbf{R})|^2 \\ \mathbf{x}_2 \sim |\Psi(\mathbf{R}')|^2}} \left[ \frac{ \Psi(\mathbf{R}_\text{A}', \mathbf{R}_\text{B}) \Psi(\mathbf{R}_\text{A}, \mathbf{R}_\text{B}') }{ \Psi(\mathbf{R}_\text{A}, \mathbf{R}_\text{B}) \Psi(\mathbf{R}_\text{A}', \mathbf{R}_\text{B}')} \right],
\end{equation}
was used for instance in \cite{2010_hastings_renyi2,2012_tubman_renyi2_vmc, 2014_herdman_renyi2_pimc,2023_sinibaldi_unbiasing,2024_mauron_topological_entanglement}, where $\mathbf{x}_1 = \mathbf{R}_\text{A} \cup \mathbf{R}_\text{B}$ and $\mathbf{x}_2 = \mathbf{R}_\text{A}' \cup \mathbf{R}_\text{B}'$, with $\mathbf{R}_\text{A}$ ($\mathbf{R}_\text{B}$) being the set of single-particle coordinates in partition A (B) (and similarly for $\mathbf{x}_2$).
However, outliers can often skew the distribution of swap estimators.

To eliminate their effect,  we use an idea from \cite{2010_van_den_nest_simulating_quantum_computers_probabilistic} to truncate the swap estimator values with the following step function
\begin{equation}
    \varepsilon(\mathbf{x},\mathbf{y}) \equiv \varepsilon(\{\mathbf{x}_1,\mathbf{x}_2\}, \{\mathbf{y}_1,\mathbf{y}_2\}) 
    = \begin{cases}
        1, & \left| \frac{ \Psi(\mathbf{y}_1) \Psi(\mathbf{y}_2) }{ \Psi(\mathbf{x}_1) \Psi(\mathbf{x}_2) } \right| < 1, \\
        0, & \text{else},
    \end{cases}
\end{equation}
and noticing that the following identity holds
\begin{equation} \label{eq:truncation_identity}
    \varepsilon(\mathbf{x},\mathbf{y}) + \varepsilon(\mathbf{y},\mathbf{x}) + \delta_{\left| \frac{ \Psi(\mathbf{y}_1) \Psi(\mathbf{y}_2) }{ \Psi(\mathbf{x}_1) \Psi(\mathbf{x}_2) } \right|, 1} = 1.
\end{equation}
Indeed, this last identity takes into account the three possible outcomes of the swap estimator: either it is (1) less than 1, (2) greater than 1 or (3) equal to 1. 
By using \cref{eq:truncation_identity}, we find that the following estimator slightly lowers the variance
\begin{align}
    \text{Tr}_\text{A} \rho_\text{A}^2 
    &= \mathbb{E} \left[ 2 \text{Re} \left[ \frac{ \Psi(\mathbf{y}_1) \Psi(\mathbf{y}_2) }{ \Psi(\mathbf{x}_1) \Psi(\mathbf{x}_2) } \right] \epsilon(\mathbf{x},\mathbf{y}) \right] \\
    & \hspace{-5mm} + \mathbb{E} \left[ e^{i \text{Arg}[ \Psi(\mathbf{y}_1) \Psi(\mathbf{y}_2) \Psi^*(\mathbf{x}_1) \Psi^*(\mathbf{x}_2)]} \delta_{\left| \frac{ \Psi(\mathbf{y}_1) \Psi(\mathbf{y}_2) }{ \Psi(\mathbf{x}_1) \Psi(\mathbf{x}_2) } \right|, 1}  \right],
\end{align}
where the probability distribution for both statistical averages is a product of two independent ground state Born distributions.

Periodic systems do not have a preferred origin.
As a result, for each Monte Carlo sample, we randomly displace the center of partition A in the simulation cell, denoted $\mathcal{C} \equiv [0,L_x] \times [0,L_y]$, to calculate different instances of the $\text{Tr}_\text{A} \rho_\text{A}^2$ estimator.
For 2D helium-4, this averaging over different origins is especially relevant in the solid phase, where the triangular lattice forms.
In particular, let $\mathbf{o} \sim \mathcal{U}(\mathcal{C})$ be the origin associated to a given Monte Carlo sample $\mathbf{x}_1 = \mathbf{R}_\text{A} \cup \mathbf{R}_\text{B}$, where $\mathcal{U}(\mathcal{C})$ denotes a uniform distribution over the simulation cell $\mathcal{C}$. 
The circle of radius $R$ and center $\mathbf{o}$ then defines partition A.
The set of single-particle coordinates in partition A (and in the Monte Carlo sample $\mathbf{x}_1$) is thus given by 
\begin{equation}
\mathbf{R}_\text{A} \equiv \left\{\mathbf{r}_i \in \mathbf{x}_1 : d_\mathrm{mic}(\mathbf{r}_i, \mathbf{o}) < R \right\},
\end{equation}
where the minimum image convention distance is defined as $d_\mathrm{mic}(\mathbf{a},\mathbf{b}) = |(\mathbf{a}-\mathbf{b}+\mathbf{L}/2) \ \text{mod} \ \mathbf{L}-\mathbf{L}/2|$, $\forall \ \mathbf{a},\mathbf{b} \in \mathcal{C}$.
Partition B is defined as the complement of partition A, i.e. $\mathbf{R}_\text{B} = \mathbf{R} \setminus \mathbf{R}_\text{A}$.
Another origin $\mathbf{o}'$ can be randomly selected for a second sample $\mathbf{x}_2 = \mathbf{R}_\text{A}' \cup \mathbf{R}_\text{B}'$.
The swapped configurations are then given by 
\begin{align}
    \mathbf{y}_1 &= (\mathbf{R}_\text{A}' - \mathbf{o}' + \mathbf{o}, \mathbf{R}_\text{B}), \\
    \mathbf{y}_2 &= (\mathbf{R}_\text{A} - \mathbf{o} + \mathbf{o}', \mathbf{R}_\text{B}').
\end{align}

\twocolumngrid
\bibliographystyle{apsrev4-2}
\bibliography{main.bib}

\clearpage
\onecolumngrid
\appendix

\section*{Supplemental Material}

\section{Helium potential energy under periodic boundary conditions} \label{appendix:potential_and_pbcs}

\begin{figure}[!b]
    \centering
    \includegraphics[width=0.6\textwidth]{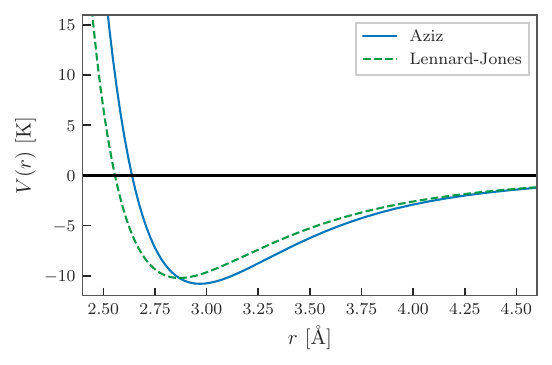}
    \caption{Comparison of the Aziz and Lennard-Jones potentials, given respectively in \cref{eq:aziz_pair_potential,eq:LJ_potential}, modeling the interatomic interaction between ${}^4$He. }
    \label{fig:aziz_vs_lennard-jones}
\end{figure}

We use the Aziz HFDHE2 pair potential \cite{1979_aziz_potential} to model the interatomic interactions between helium atoms.
The electrons are thus abstracted away while the helium atoms are treated as elementary particles.
For completeness, we write the Aziz potential explicitly here
\begin{equation} \label{eq:aziz_pair_potential}
    v_\text{Aziz}(x) = \epsilon \left[ A e^{-\alpha x} - \left( \frac{C_6}{x^6} + \frac{C_8}{x^8} + \frac{C_{10}}{x^{10}}
 \right) F(x) \right], 
 \qquad F(x) = 
 \begin{cases}
 e^{-(D/x-1)^2}, & x < D, \\
 1, & x \ge D,
 \end{cases}
\end{equation}
where $x = r/r_m$, with $r$ being the distance between two helium atoms, $r_m = 2.9673$ {\AA} is the minimum of the potential, and with the parameters
\begin{equation}
    \begin{cases}
    \epsilon/k_B = 10.8 \ \text{K}, \\
    A = 0.544850 \times 10^6, \\
    \alpha = 13.353384, \\
    C_6 = 1.3732412, \\
    C_8 = 0.4253785, \\
    C_{10} = 0.178100, \\
    D = 1.241314.
\end{cases}
\end{equation}
This potential is shown in \cref{fig:aziz_vs_lennard-jones}.
One can see that it qualitatively resembles the Lennard-Jones potential
\begin{equation} \label{eq:LJ_potential}
v_\text{LJ}(r) = 4 \tilde{\epsilon} \left[ \left( 
\frac{\sigma}{r} \right)^{12} - \left( \frac{\sigma}{r} \right)^6 \right],
\end{equation}
with $\Tilde{\epsilon} = 10.22$ K and $\sigma = 2.556$ \AA.
In fact, prior to the work of Aziz \textit{et al.} \cite{1979_aziz_potential}, the Lennard-Jones potential was used to describe the interactions between helium atoms, for instance in \cite{1954_kilpatrick_lennard-jones,1965_mcmillan_ground_state_he4,1976_liu_lennard-jones_he4}.

Here we impose periodic boundary conditions (PBCs) to investigate the bulk properties of the system.
It is also natural to impose PBCs here since a triangular lattice - an inherently periodic structure - forms under sufficiently large pressures (or equivalently large number densities).
We now present two different functional forms for the potential energy $V(\mathbf{R})$.

\subsection{Periodized potential} \label{appendix:periodized_potential}

\begin{figure}[h]
    \centering
    \includegraphics[width=0.7\textwidth]{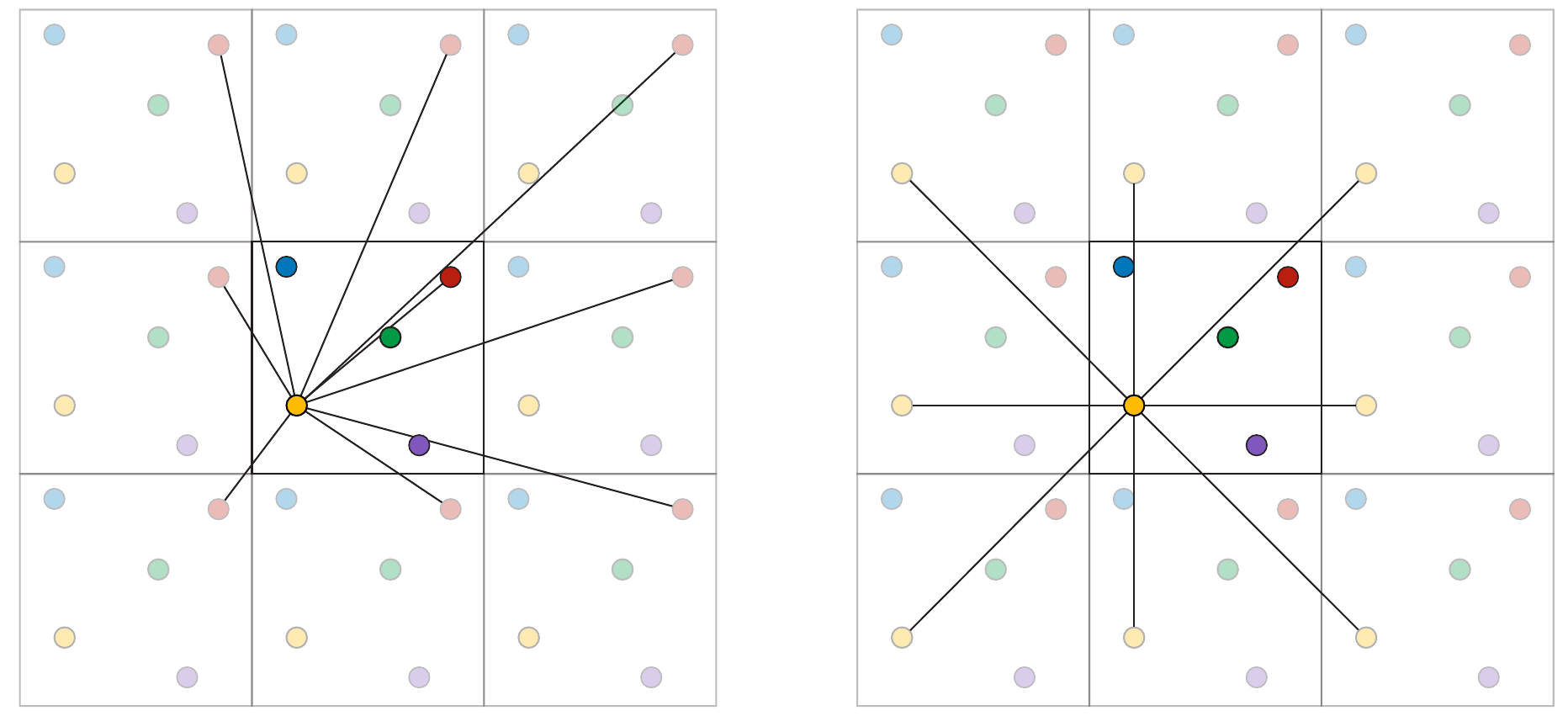}
    \caption{
    The two types of interactions appearing in the potential energy given in \cref{eq:splitted_periodized_potential} are depicted: (1) interactions with different particles (on the left) and (2) self-interactions (on the right).
    For simplicity, $N=5$ particles are shown in a square simulation cell (in bold).
    The continuous space $\mathbb{R}^2$ is tiled with identical copies of that simulation cell as PBCs are imposed.
    }
    \label{fig:potential_contributions_sketch}
\end{figure}

The full periodized potential reads
\begin{equation} \label{eq:periodized_potential_energy}
    V(\mathbf{R}) = \frac{1}{2} \sum_{i,j} \sum_{\mathbf{n}} {}^{'} v(|\mathbf{r}_{ij} + \mathbf{n} \mathbf{L}|),
\end{equation}
where $v(r) \equiv v_\text{Aziz}(r/r_m)$, $\mathbf{r}_{ij}=\mathbf{r}_i-\mathbf{r}_j$ is the distance vector between particle $i$ and particle $j$, $\mathbf{L}=(L_x,L_y)$ is the size of the simulation cell, and the restriction on the sum specifies that terms where $i=j$ and $\mathbf{n}=\mathbf{0}$ should not be included. 
It is then possible to rewrite this sum in the following way
\begin{align}
    V(\mathbf{R}) 
    &= \frac{1}{2} \sum_{i,j} \sum_\mathbf{n} v(|\mathbf{r}_{ij} + \mathbf{n} \mathbf{L}|) - \frac{1}{2} \sum_{i=j} v(|\mathbf{r}_{ij}|) \\
    &= \frac{1}{2} \sum_{i,j} \sum_\mathbf{n} v(|\mathbf{r}_{ij} + \mathbf{n} \mathbf{L}|) - \frac{1}{2} \sum_{i=j} \sum_\mathbf{n} v(|\mathbf{r}_{ij} + \mathbf{n} \mathbf{L}|) + \frac{1}{2} \sum_{i=j} \sum_{\mathbf{n} \ne \mathbf{0}} v(|\mathbf{r}_{ij} + \mathbf{n} \mathbf{L}|) \\
    &= \sum_{i<j} \sum_{\mathbf{n}} v(|\mathbf{r}_{ij} + \mathbf{n} \mathbf{L}|) + \frac{N}{2} \sum_{\mathbf{n} \ne \mathbf{0}} v(| \mathbf{n} \mathbf{L}|), \label{eq:splitted_periodized_potential}
\end{align}
where the second term corresponds to self-interactions, and is referred to as the Madelung constant.
In practice, a finite number of images are taken into account by fixing a cutoff on the sum over the grid points $\mathbf{n}$.
A sketch of the periodized simulation cell with each potential term in \cref{eq:splitted_periodized_potential} is shown in \cref{fig:potential_contributions_sketch}.

\subsection{Shifted and truncated potential} \label{sec:shifted_and_truncated_potential}

Instead of considering all periodic images (up to a chosen cutoff), as in \cref{appendix:periodized_potential}, the potential can be restricted to account for only a single image, provided corrections are taken into account.
In this case, the full potential energy $V(\mathbf{R})$ can be written as follows
\begin{equation} \label{eq:full_shifted_and_truncated_potential}
    V(\mathbf{R}) = V_s + V_t + \sum_{i < j} \Tilde{v}(r_{ij}),
\end{equation}
where only the third term depends on the Monte Carlo samples.
It is written in terms of a shifted and truncated potential of the form
\begin{equation} \label{eq:shifted_and_truncated_potential}
    \Tilde{v}(r) = 
    \begin{cases}
        v(r) - v(r_c), & r \le r_c, \\
        0, &  \text{else},
    \end{cases}
\end{equation}
where $v(r) \equiv v_\text{Aziz}(r/r_m)$ and $r_c = \min(\mathbf{L})/2$ is used as a cutoff distance to consider only a single image.
The distances fed to the potential $\Tilde{v}(r)$, given in \cref{eq:shifted_and_truncated_potential}, are calculated using the minimum image convention distance
\begin{equation} \label{eq:minimum_image_convention_distance}
    d_\text{mic}(\mathbf{r}_i, \mathbf{r}_j) = \left| \left(\mathbf{r}_i - \mathbf{r}_j + \frac{\mathbf{L}}{2} \right) \text{mod} \ \mathbf{L} - \frac{\mathbf{L}}{2} \right|.
\end{equation}
To account for the fact that the potential was shifted and truncated, the shift and tail contributions, respectively $V_s$ and $V_t$, were introduced in \cref{eq:full_shifted_and_truncated_potential}.
They are independent of the Monte Carlo samples, and are extensive quantities defined as \cite{2002_frenkel_molecular_simulation}
\begin{align}
    V_s &= N \frac{N_c}{2} v(r_c),\\
    V_t &= N \frac{n}{2} \int \dd^d r \theta(r-r_c) v(r),
\end{align}
where $\theta$ is the step function.
The number of particles in the sphere of radius $r_c$, denoted $N_c$, is given by $N_c = V_d(r_c) n$, with $V_d(r) = \pi^{d/2} r^d/ \Gamma(d/2+1)$ being the volume of a $d$-dimensional sphere of radius $r$, and $n=N/V$ is the number density. 
The tail contribution can be written more explicitly as follows
\begin{equation}
    V_t = N n \frac{ \pi^{d/2} }{ \Gamma \left( \frac{d}{2} \right) } \int_{r_c}^\infty dr r^{d-1} v(r),
\end{equation}
where the solid angle subtended by a $d$-dimensional sphere $\Omega_d = 2 \pi^{d/2}/\Gamma(d/2)$ results from the angular integration.

\section{Helium-4 ground state trial wavefunction} \label{appendix:helium_ground_state_implementation}

\textbf{Ansatz:}
Let $\mathbf{R}=(\mathbf{r}_1,\hdots,\mathbf{r}_N) \in \mathbb{R}^{N \times d}$ denotes all single-particle coordinates.
To determine a proper trial wavefunction, we proceed similarly as in \cite{2024_pescia_mpnn_electrongas} and use the fact that the exact ground state $\Psi_0(\mathbf{R})$ can be obtained by taking the limit of large imaginary times of $\Phi_\tau(\mathbf{R}) \equiv \langle \mathbf{R} | e^{-\tau H} | \Phi \rangle$, that is $\Psi_0(\mathbf{R}) \propto \lim_{\tau \to \infty} \Phi_\tau(\mathbf{R})$, provided that the chosen reference state $\Phi(\mathbf{R})$ satisfies $\langle \Phi | \Psi_0 \rangle \neq 0$. 
By adding a completeness relation, the propagation of the reference state in imaginary times reads
\begin{equation} \label{eq:imag_evol_ref_state}
    \Phi_\tau(\mathbf{R}) = \int d \mathbf{R}' G_\tau(\mathbf{R}, \mathbf{R}') \Phi(\mathbf{R}').
\end{equation}
We construct the variational trial wavefunction $\Psi_{\boldsymbol{\theta}}(\mathbf{R})$ by modeling the propagator $G_\tau(\mathbf{R}, \mathbf{R}')$ with backflowed coordinates fed to a multi-layer perceptron (MLP), and the reference state $\Phi(\mathbf{R}')$ with the McMillan's ansatz \cite{1965_mcmillan_ground_state_he4}. 
This bosonic trial wavefunction is given by
\begin{equation} \label{eq:trial wavefunction}
    \Psi_{\boldsymbol{\theta}}(\mathbf{R}) = \prod_{i<j} \exp[\theta_1 \text{MLP}(\mathbf{e}_{ij}^{(K)})] \exp \left[ - \frac{\theta_2}{d_\text{sin}(\mathbf{r}_i, \mathbf{r}_j)^{5}} \right],
\end{equation}
where the mean value theorem for integrals was used to abstract away the integral in \cref{eq:imag_evol_ref_state}, as motivated in \cite{2024_pescia_mpnn_electrongas}.
Only two variational parameters $\theta_1, \theta_2 \in \mathbb{R}$ are explicitly shown, while others are hidden inside the MLP and the backflowed coordinates $\mathbf{e}_{ij}^{(K)}$.
These coordinates are the output of a graph neural network (GNN) after $K$ message passing iterations.
The backflowed coordinates are symmetrized via the transformation $\mathbf{e}^{(K)} \mapsto (\mathbf{e}^{(K)} + (\mathbf{e}^{(K)})^\text{T})/2$ before the MLP processing (with the transpose acting on the particle indices).
The sine distance $d_\text{sin}(\mathbf{r}_i, \mathbf{r}_j) = |(\mathbf{L}/2) \sin (\pi \mathbf{r}_{ij}/ \mathbf{L})|$ was further introduced and $\mathbf{r}_{ij} = \mathbf{r}_i - \mathbf{r}_j$.
These operations ensure that the trial wavefunction is invariant under the permutation of single-particle coordinates, and the sine distance additionally imposes the PBCs.

For completeness, we now provide details on the GNN implementation used, even though details were previously given for instance in \cite{2024_pescia_mpnn_electrongas,2024_kim_ultra_cold_fermi_gases,2024_nys_tdvmc}.
The implementation is based on the message passing scheme \cite{2017_gilmer_mpnn_for_chemistry}, as well as the attention mechanism \cite{2017_google_self_attention}.
Other works developed and refined these ideas \cite{2018_velickovic_graph_attention_networks, 2018_battaglia_graph_networks}. 

\textbf{Graph structure:} 
Let $\mathcal{G} = (\mathcal{V},\mathcal{E})$ denote an undirected graph characterized by a set of vertices $\mathcal{V}$ and a set of edges $\mathcal{E}$.
We construct $\mathcal{V}$ by a one-to-one mapping of the particle index: $\{1,\hdots,N\} \mapsto \{v_1, \hdots, v_N\} \equiv \mathcal{V}$.
While each vertex $v_i \in \mathcal{V}$ is associated to an atom, an edge $e_{ij} \in \mathcal{E}$ in turn models the pairwise interactions between atoms.
We only consider complete graphs $\mathcal{G}$ to avoid unnecessary assumptions on the adjacency of vertices.

\textbf{Vertex and edge features:} 
For each vertex $v_i \in \mathcal{V}$ and edge $e_{ij} \in \mathcal{E}$, we associate feature vectors $\mathbf{v}_i^{(k)}$ and $\mathbf{e}_{ij}^{(k)}$, respectively, where the additional index $k$ specifies the message passing iteration (see below). 
The vertex and edge feature vectors are both initialized with hidden variational parameters, denoted respectively by $\mathbf{h}_i \in \mathbb{R}^{n_f}$ and $\mathbf{H}_{ij} \in \mathbb{R}^{n_f}$, where $n_f$ is a chosen number of features.
The edge feature vector is supplemented with the (static) physical information of the problem.
This information is packaged in the input tensor
\begin{equation}
\mathbf{I}_{ij} \equiv [\sin(2\pi \mathbf{r}_{ij}/\mathbf{L}), \cos(2\pi \mathbf{r}_{ij}/\mathbf{L}), |\sin(\pi \mathbf{r}_{ij}/\mathbf{L})|] \in \mathbb{R}^{2d+1},
\end{equation}
which correspond to periodized distance coordinates and periodized distances, and the square brackets correspond to the concatenation operation.
The initialization $(k=0)$ of the feature vectors is therefore
\begin{align}
    \mathbf{v}_i^{(0)} &= \mathbf{h}_i \in \mathbb{R}^{n_f}, \\
    \mathbf{e}_{ij}^{(0)} &= [\mathbf{I}_{ij},\mathbf{H}_{ij}] \in \mathbb{R}^{2d+1+n_f}.
\end{align}

\textbf{Particle-attention-based message passing:} 
At every message passing iteration $k$, query and key matrices $Q^{(k)},K^{(k)} \in \mathbb{R}^{n_a \times \text{dim}(\mathbf{e}_{ij}^{(k)})}$ are initialized, where $n_a$ is the number of so-called attention heads \cite{2017_google_self_attention}.
For every pair of particles $(i,j)$, query and key vectors are then calculated using the same query and key matrices: $\mathbf{Q}_{ij}^{(k)} = Q^{(k)} \mathbf{e}_{ij}^{(k)}, \ \ \mathbf{K}_{ij}^{(k)} = K^{(k)} \mathbf{e}_{ij}^{(k)} \in \mathbb{R}^{n_a}$.
Similarly as in \cite{2024_pescia_mpnn_electrongas}, rather than using a linear map (as for the query and key vectors), we use a multi-layer perceptron (MLP) to compute the so-called values, that is, $\mathbf{V}_{ij}^{(k)} = \text{MLP}_\mathbf{V}^{(k)}(\mathbf{e}_{ij}^{(k)}) \in \mathbb{R}^{n_e}$.
In this work, MLPs always operate on the feature dimension and they output vectors of $n_e$ embedding entries.
We label the MLPs with different subscripts to specify that they do not share the same weights and biases.
The attention mechanism can then be used to compute messages
\begin{equation}
    \mathbf{M}_{ij}^{(k)} = \mathbf{W}_{ij}^{(k)} \odot \mathbf{V}_{ij}^{(k)} \equiv \text{MLP}_\mathbf{W}^{(k)} \left( \frac{ \sum_{l=1}^N \mathbf{Q}_{il}^{(k)} \mathbf{K}_{lj}^{(k)} }{\sqrt{n_a}} \right) \odot \mathbf{V}_{ij}^{(k)},
\end{equation}
where we implicitly defined the attention weights vectors $\mathbf{W}_{ij}^{(k)} \in \mathbb{R}^{n_e}$, and used the symbol $\odot$ to denote element-wise multiplication.
We did not find a benefit to converting the attention weights to probabilities by applying a softmax as in the original self-attention paper \cite{2017_google_self_attention}.
We aggregate all the messages arriving at the vertex $i$ with a permutation invariant pooling function, taken by default to be a sum, so that
\begin{equation}
    \mathbf{m}_i^{(k)} = \sum_{j=1}^N \mathbf{M}_{ij}^{(k)}.
\end{equation}
When increasing $N$, it is eventually necessary to normalize the sum to avoid numerical overflows.
This can be achieved by taking the mean instead of a plain sum.
In principle, mean pooling could also facilitate the transfer learning process for different $N$, though this was not explored in this work.
Moreover, if the underlying graph $\mathcal{G}$ was not assumed to be complete, the sum over the messages would be restricted to vertices $j$ in a neighborhood $\mathcal{N}(i)$ of the vertex $i$ (and not the entire set of vertices $\mathcal{V}$).

\textbf{Updating vertex and edge features:}
We update the vertex and edge feature vectors using the messages, and embed them to another latent space using MLPs via
\begin{align}
    \mathbf{v}_i^{(k+1)} &= \text{MLP}_\mathbf{v}^{(k)}([\mathbf{v}_i^{(k)}, \mathbf{m}_i^{(k)}]) \in \mathbb{R}^{n_e}, \\
    \mathbf{e}_{ij}^{(k+1)} &= [\mathbf{I}_{ij}, \text{MLP}_\mathbf{e}^{(k)}([\mathbf{e}_{ij}^{(k)}, \mathbf{M}_{ij}^{(k)}])] \in \mathbb{R}^{2d+1+n_e}.
\end{align}
The input tensor $\mathbf{I}_{ij}$ is concatenated with the edge feature vectors as a skip connection to avoid erasing the physical information after a few message passing iterations.
The message passing scheme can be iterated $K$ times, at which point the final vertex and edge feature vectors $\mathbf{v}_i^{(K)}$ and $\mathbf{e}_{ij}^{(K)}$ are returned for all $1 \le i,j \le N$.
Notice however that in the trial wavefunction given in \cref{eq:trial wavefunction}, only the edge features are used.

\section{Variational and Diffusion Monte Carlo energy comparison}
To assess the accuracy of the NQS trial wavefunction used in this work, we compare in \cref{table:vmc_vs_dmc} the associated variational Monte Carlo (VMC) energies to those obtained with a wavefunction of the form
\begin{equation} \label{eq:u2_u3_ansatze}
    \log \Psi_{U2/3}(\mathbf{R}) = -\sum_{i<j} u_2(r_{ij}) - \sum_i \left[\sum_j \mathbf{r}_{ij} u_3(r_{ij}) \right] 
    \cdot \left[\sum_j \mathbf{r}_{ij} \widetilde{u}_3(r_{ij})
    \right].
\end{equation}
To simulate the solid phase, $n \gtrsim 0.070$ \AA$^{-2}$,
additional single-body Gaussians are used to localize the particles at the triangular lattice sites.
We denote the case where only two-body correlations are included as VMC U2, in which ase $u_3 = 0$.
The case VMC U3 then corresponds to the more general case where $u_2$, $u_3$, and $\widetilde{u}_3$ are optimized.
Although diffusion Monte Carlo (DMC) energies eventually converge to the true ground state energy, in practice, this can be difficult if the trial wavefunction does not have a significant overlap with the true ground state \cite{2012_boninsegni_population_size_bias_dmc}.

From \cref{table:vmc_vs_dmc}, we see the importance of the many-body correlations included in the NQS to reach DMC level accuracy.
In particular, the variance per particle $\sigma^2/N$ of the NQS wavefunction is decreased by up to two orders of magnitude compared to VMC U3.
The NQS energies remain slighly above the DMC energies only for $N=80$ at the solid densities $n=0.070$ \AA$^{-2}$ and $n=0.075$ \AA$^{-2}$.

\begin{table}[!h]
    \centering
    \renewcommand{\arraystretch}{1.5} 
    \setlength{\tabcolsep}{8pt} 
    \begin{tabular}{|c|c|c|c|c|c|c|c|c|}
    \cline{3-9}
    \multicolumn{2}{c}{} & \multicolumn{2}{|c}{VMC U2} & \multicolumn{2}{|c|}{VMC U3} & DMC & \multicolumn{2}{|c|}{VMC NQS} \\
    \hline
    $N$ & $n$ [\AA$^{-2}$] & $E/N$ & $\sigma^2/N$ & $E/N$ & $\sigma^2/N$ & $E/N$ & $E/N$ & $\sigma^2/N$
    \\
    \hline \hline
    \multirow{5}{*}{30} & 0.065 & 0.632(5) & 45.0(5) & 0.132(3) & 17.8(1) & -0.047(1) & -0.046(1) & 0.11 \\
    & 0.068 & 1.073(8) & 55.3(9) &  0.477(4) & 9.7(1) & 0.245(2) & 0.246(2) & 0.25 \\  
    & 0.070 & 0.684(4) & 14.9(5) & 0.581(2) & 4.28(7) & 0.461(2) & 0.450(2) & 0.31 \\
    & 0.072 & 0.867(4) & 16.1(2)  & 0.765(2) & 4.52(7) & 0.661(2) & 0.659(2) & 0.29  \\
    & 0.075 & 1.226(3) & 19.9(3) & 1.101(2) & 4.49(8) & 1.020(1) & 1.020(1) & 0.24  \\
    \hline
    \multirow{4}{*}{80} & 0.065 & 0.667(6) & 43.1(1) & 0.268(4) & 15.1(2) & -0.021(1) & -0.020(1) & 0.15 \\
    & 0.068 & 1.133(8) & 53(1) &  0.640(4) & 18.9(2) & 0.287(1) & 0.290(1) & 0.20 \\  
    &0.070 & 0.738(3) & 14.8(2) & 0.650(2) & 4.52(7) & 0.504(2) & 0.518(3) & 1.17 \\
    & 0.075 & 1.305(3) & 18.1(2) & 1.195(2) & 4.08(4) & 1.073(1) & 1.085(3) & 0.97 \\
    \hline
    \end{tabular}
    \caption{
    Ground state energies using a truncated and shifted version of the Aziz potential (described in \cref{sec:shifted_and_truncated_potential}).
    Energies (per particle) $E/N$ are in units of kelvin while energy variances (per particle) $\sigma^2/N$ are in units of kelvin squared.
    The neural quantum state results (VMC NQS) obtained in this work are compared with two simpler wavefunctions, including
   two-body pair correlations (VMC U2), as well as additional three-body correlations (VMC U3) (see \cref{eq:u2_u3_ansatze}). 
   In the solid phase, $n \gtrsim 0.070$ \AA$^{-2}$, the VMC U2 and VMC U3 trial wavefunctions contain additional gaussians to explicitly impose a triangular crystal lattice. 
   The stochastically improved Diffusion Monte Carlo values (DMC)
   are based on VMC U3.
   }
   \label{table:vmc_vs_dmc}
\end{table}
\section{Maxwell construction} \label{appendix:maxwell_construction}

\begin{figure}[!t]
    \centering
    \includegraphics[width=0.6\textwidth]{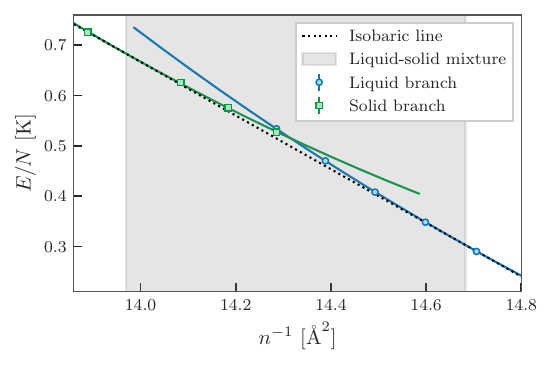}
    \caption{
    The liquid-solid coexistence region (shaded in grey), obtained with the common tangent (dotted line) of the Maxwell construction, is depicted for $N=80$ ${}^4$He atoms.
    The freezing and melting number densities found, by numerically solving the set of equations given in \cref{eq:maxwell_construction_equations}, are respectively $n_f=0.0681$ \AA$^{-2}$ and $n_m=0.0716$ \AA$^{-2}$. 
    The coexistence region is defined by the interval $[n_m^{-1},n_f^{-1}]$.
    The associated critical pressure, defined in \cref{eq:critical_pressure}, is $P_c = 0.53(1)$ K\AA$^{-2}$.
    }
    \label{fig:phase_diagram_N=80_canonical_ensemble}
\end{figure}

While in the main text no metastable liquid and solid branches were found for $N=30$, these branches can eventually arise if the free energy barrier between the liquid and solid states is getting too large.
Indeed, this is shown in the main text in Fig. 1 as an inset for $N=80$, and the same data points are shown in \cref{fig:phase_diagram_N=80_canonical_ensemble} here.
In this case, the relaxation time between the two symmetry sectors can be extremely long, making it very challenging to recover the common tangent line between the liquid and solid phases.
In this section, we discuss how to systematically determine this common tangent (isobaric) line using the Maxwell construction \cite{1998_gordillo_phase_diagram_he4, 2000_krishnamachari_shadow_wavefunction}. 
This approach will also enable us to estimate the freezing and melting densities.

The metastable liquid (solid) branch shown in \cref{fig:phase_diagram_N=80_canonical_ensemble} was obtained by transfer learning, using the variational parameters optimized at number densities deeper in the liquid (solid) phase.
The thermodynamic equilibrium is depicted by the isobaric dotted line in \cref{fig:phase_diagram_N=80_canonical_ensemble}, which corresponds to the common tangent line.
This line can be systematically found by solving numerically the following system of two equations with respect to the freezing and melting number densities, $n_f$ and $n_m$ respectively,
\begin{equation} \label{eq:maxwell_construction_equations}
\begin{cases}
    l'(n_f^{-1}) &= s'(n_m^{-1}), \\
    l'(n_f^{-1}) &= \dfrac{s(n_m^{-1}) - l(n_f^{-1})}{n_m^{-1} - n_f^{-1}},
\end{cases}
\end{equation}
where $l$ and $s$ are the liquid and solid branch functional form, taken here to be (quartic) spline fits.
Given that the thermodynamic pressure is defined as $P = - \partial E/ \partial V$, the critical pressure associated with the common tangent line can be approximated with the following expression
\begin{equation} \label{eq:critical_pressure}
    P_c \approx - \frac{s(n_m^{-1}) - l(n_f^{-1})}{n_m^{-1}-n_f^{-1}}.
\end{equation}
The critical pressure of the common tangent line shown in \cref{fig:phase_diagram_N=80_canonical_ensemble} is $P_c = 0.53(1)$ K\AA$^{-2}$.
This result is comparable in magnitude to the critical pressures found at finite temperatures in \cite{1998_gordillo_phase_diagram_he4}.

\section{Finite size corrections of the total energy and pressure}

Despite the use of periodic boundary conditions, calculations are done in a finite simulation box containing $N$ atoms, and the corresponding energy per particle, $\varepsilon_N = E_N/N$, needs to be extrapolated to the thermodynamic limit value, $\varepsilon_\infty=\delta E_N/N + \delta \varepsilon_N$. 

\begin{figure}[!h]
  \centering
  \includegraphics[width=\textwidth]{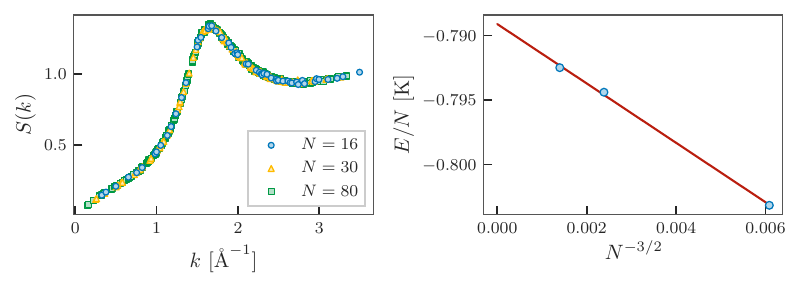}
  \caption{
  Finite size effects at a (liquid) number density of $n=0.05$ \AA$^{-2}$. (Left) The static structure factor is shown for three different number of particles $N$.
  (Right) The ground state energy per particle $E/N$ is extrapolated to the thermodynamic limit by taking into account the $\sim N^{-3/2}$ energy correction obtained using the framework presented in \cite{2016_holzmann_finite_size_effects}.
  The red line thus corresponds to the fit $f(N)=a N^{-3/2} + b$, for some fitting parameters $a$ and $b$.
  For the liquid density of interest, the intercept is $b \approx -0.789$ K, which corresponds to the extrapolated energy $\varepsilon_\infty \equiv E_\infty/N$.
  The calculations were performed using the shifted and truncated potential, as described \cref{sec:shifted_and_truncated_potential}, in the canonical ensemble.
  }
  \label{fig:finite_size_effects_panel_plot}
\end{figure}

As discussed in \cite{2006_chiesa_finite_size_error}, leading order size effects are related to the infrared behavior of the structure factor $S(k \to 0)$.
In \cref{fig:finite_size_effects_panel_plot} (Left), we plot $S(k)$ for different system sizes, $N=16$, $30$ and $80$, and observe that they lie on top of each other independently of $N$. 
Our data also confirms the expected linear behavior $S(k) \sim k$ for small $k$ \cite{feenberg2012theory}, which implies finite size corrections of the form $\delta \varepsilon_N \sim N^{-3/2}$  \cite{2016_holzmann_finite_size_effects}.
In \cref{fig:finite_size_effects_panel_plot} (Right), we display an energy extrapolation line scaling like $\sim N^{-3/2}$ for the liquid density $n=0.05$ \AA$^{-2}$.

With the non-extrapolated energies $\varepsilon_N$, we obtain the critical pressures $P_{c,N=30}=0.50(1)$ K\AA$^{-2}$, $P_{c,N=56}=0.53(1)$ K\AA$^{-2}$ and $P_{c,N=80}=0.53(1)$ K\AA$^{-2}$ using \cref{eq:critical_pressure} with $n=0.068$ \AA$^{-2}$ and $n=0.072$ \AA$^{-2}$ (corresponding to the data points closest to the freezing and melting number densities approximated in \cref{appendix:maxwell_construction}).
With the numerically extrapolated energies $\varepsilon_\infty$ as a function of number density, we obtain $P_{c,\infty}=0.54(1)$ K\AA$^{-2}$ as a thermodynamic estimate for the critical pressure.

\section{Gradient estimator} \label{appendix:gradient_estimator}

Consider an operator $A(\boldsymbol{\theta})$ and a trial wavefunction $\Psi_{\boldsymbol{\theta}}(\mathbf{R})$ that both depend on the variational parameters $\boldsymbol{\theta}$.
We briefly drop the dependence of $\Psi_{\boldsymbol{\theta}}$ on the coordinates $\mathbf{R}$ to simplify the notation. 
We want to differentiate the expectation value $\langle A(\boldsymbol{\theta}) \rangle$ with respect to the variational parameters $\boldsymbol{\theta}$, that is, 
\begin{equation} \label{eq:gradient_of_arbitrary_operator}
    \nabla_{\boldsymbol{\theta}} \frac{\langle \Psi_{\boldsymbol{\theta}} | A(\boldsymbol{\theta}) | \Psi_{\boldsymbol{\theta}} \rangle}{\langle \Psi_{\boldsymbol{\theta}} | \Psi_{\boldsymbol{\theta}} \rangle}
    =  \frac{ \int d\mathbf{R} [ \nabla_{\boldsymbol{\theta}} \Psi_{\boldsymbol{\theta}}^* ] A(\boldsymbol{\theta}) \Psi_{\boldsymbol{\theta}} }{ \int d\mathbf{R} |\Psi_{\boldsymbol{\theta}} |^2 }
    +
    \frac{ \int d\mathbf{R} \Psi_{\boldsymbol{\theta}}^* A(\boldsymbol{\theta}) [\nabla_{\boldsymbol{\theta}} \Psi_{\boldsymbol{\theta}}] }{ \int d\mathbf{R} |\Psi_{\boldsymbol{\theta}} |^2 }
    +
    \frac{ \int d\mathbf{R} \Psi_{\boldsymbol{\theta}}^* A(\boldsymbol{\theta}) \Psi_{\boldsymbol{\theta}} }{ \nabla_{\boldsymbol{\theta}}[\int d\mathbf{R} |\Psi_{\boldsymbol{\theta}} |^2] }
    + \frac{\int d\mathbf{R} \Psi_{\boldsymbol{\theta}}^* [\nabla_{\boldsymbol{\theta}} A(\boldsymbol{\theta})] \Psi_{\boldsymbol{\theta}}}{\int d\mathbf{R}|\Psi_{\boldsymbol{\theta}}|^2},
\end{equation}
where the last term can already be recognized as $\langle \nabla_{\boldsymbol{\theta}} A(\boldsymbol{\theta}) \rangle = \mathbb{E}[\nabla_{\boldsymbol{\theta}} A_\text{loc}(\mathbf{R})]$ (dropping the dependence of the local estimator on $\boldsymbol{\theta}$ and re-introducing the dependence on the sampled coordinates $\mathbf{R}$).
As we want to work with log-amplitudes $\ln \Psi_{\boldsymbol{\theta}}$, it is convenient to rewrite the numerator of the first term of \cref{eq:gradient_of_arbitrary_operator} as follows
\begin{equation}
    [ \nabla_{\boldsymbol{\theta}} \Psi_{\boldsymbol{\theta}}^* ] H \Psi_{\boldsymbol{\theta}} 
    = \underbrace{
    [ \nabla_{\boldsymbol{\theta}} \Psi_{\boldsymbol{\theta}}^* ] \Psi_{\boldsymbol{\theta}}^{*-1} }_{= \nabla_{\boldsymbol{\theta}} \ln \Psi_{\boldsymbol{\theta}}^*
    }
    \underbrace{
    \Psi_{\boldsymbol{\theta}}^* \Psi_{\boldsymbol{\theta}}}_{
    |\Psi_{\boldsymbol{\theta}}|^2
    }
    \underbrace{
    \Psi_{\boldsymbol{\theta}}^{-1} A(\boldsymbol{\theta}) \Psi_{\boldsymbol{\theta}} }_{
    A_\mathrm{loc}
    }
    ,
\end{equation}
and similarly for the numerator of the second term.
The third term of \cref{eq:gradient_of_arbitrary_operator} simplifies to
\begin{align}
        \int d\mathbf{R} \Psi_{\boldsymbol{\theta}}^* A(\boldsymbol{\theta}) \Psi_{\boldsymbol{\theta}} \nabla_{\boldsymbol{\theta}} \left[\int d\mathbf{R}' |\Psi_{\boldsymbol{\theta}} |^2 \right]^{-1}
        &=
        - \int d\mathbf{R} \Psi_{\boldsymbol{\theta}}^* A(\boldsymbol{\theta}) \Psi_{\boldsymbol{\theta}} \frac{ \int d\mathbf{R}' \{ [\nabla_{\boldsymbol{\theta}} \Psi_{\boldsymbol{\theta}}^*] \Psi_{\boldsymbol{\theta}} + \Psi_{\boldsymbol{\theta}}^* [\nabla_{\boldsymbol{\theta}} \Psi_{\boldsymbol{\theta}}] \} }{ ( \int d\mathbf{R} | \Psi_{\boldsymbol{\theta}} |^2 )^2 } \nonumber \\
        &= 
        - \langle A(\boldsymbol{\theta}) \rangle 
        \frac{ \int d\mathbf{R} | \Psi_{\boldsymbol{\theta}} |^2 ( \nabla_{\boldsymbol{\theta}} \ln \Psi_{\boldsymbol{\theta}}^* + \nabla_{\boldsymbol{\theta}} \ln \Psi_{\boldsymbol{\theta}} ) }{ \int d\mathbf{R} | \Psi_{\boldsymbol{\theta}} |^2 }
        .
\end{align}
Putting everything together, and re-introducing the dependence on $\mathbf{R}$, we obtain the five following terms
\begin{equation}
    \nabla_{\boldsymbol{\theta}} \langle A(\boldsymbol{\theta}) \rangle 
    =
    \mathbb{E}[ E_\mathrm{loc}(\mathbf{R}) \nabla_{\boldsymbol{\theta}} \ln \Psi_{\boldsymbol{\theta}}^*(\mathbf{R}) ] - \mathbb{E}[ E_\mathrm{loc}(\mathbf{R}) ] \mathbb{E}[ \nabla_{\boldsymbol{\theta}} \ln \Psi_{\boldsymbol{\theta}}^*(\mathbf{R}) ] + \ \mathrm{c.c.} + \mathbb{E}[\nabla_{\boldsymbol{\theta}} A_\text{loc}(\mathbf{R})],
\end{equation}
where c.c. stands for the complex conjugate (of the first two terms). 
The first four terms can be grouped into a so-called force term
\begin{equation} 
    \boldsymbol{f}(\mathbf{R}) = 2 \mathrm{Re} \{ (E_\mathrm{loc}(\mathbf{R}) - \mathbb{E}[E_\mathrm{loc}(\mathbf{R})]) \nabla_{\boldsymbol{\theta}} \ln \Psi_{\boldsymbol{\theta}} (\mathbf{R})\}.
\end{equation}
The gradient entries can therefore be compactly written as 
\begin{equation} \label{eq:gradient_wrt_a_var_param}
    \frac{\partial \langle A(\boldsymbol{\theta}) \rangle}{\partial \theta_i} 
    = \frac{\partial}{\partial \theta_i} \frac{ \langle \Psi_{\boldsymbol{\theta}} | A(\boldsymbol{\theta}) | \Psi_{\boldsymbol{\theta}} \rangle }{ \langle \Psi_{\boldsymbol{\theta}} | \Psi_{\boldsymbol{\theta}} \rangle}
    = 2 \text{Re}[f_i] + \mathbb{E}[\partial_{\theta_i} A_\text{loc}(\boldsymbol{\theta})],
\end{equation}
where the force term is given by a covariance
\begin{equation}
    f_i = \text{Cov}[O_i, A_\text{loc}(\boldsymbol{\theta})] = \mathbb{E}[O_i^*(A_\text{loc}(\boldsymbol{\theta})-\mathbb{E}[A_\text{loc}(\boldsymbol{\theta})])],
\end{equation}
with $O_i(\mathbf{R}) = \partial_{\theta_i} \ln \Psi_{\boldsymbol{\theta}}(\mathbf{R})$.

\section{Isobaric ensemble simulations} \label{appendix:kinetic_estimator_isobaric_ensemble}

\textbf{Loss function:}
In the isobaric ensemble, the pressure is kept fixed while the size and the shape of the system are allowed to vary. 
Equilibrium is reached when the Gibbs free energy is minimized.
For a system of $N$ particles of mass $m$ confined in a parallelogram simulation cell, the Gibbs free energy operator is given by
\begin{equation}
    G = H + P V = \frac{-\hbar^2}{2m} \sum_{i=1}^N \left( \frac{\partial^2}{\partial x_i^2} + \frac{\partial^2}{\partial y_i^2} \right) + V(\mathbf{R}) + P L_x L_y \sin \theta,
\end{equation}
where $H$ is the Hamiltonian, $\mathbf{R} = (\mathbf{r}_1,\hdots,\mathbf{r}_N)$ denotes all $N$ single-particle coordinates $\mathbf{r}_i = (x_i,y_i)$, and $V(\mathbf{R})$ is the full potential energy described in \cref{appendix:potential_and_pbcs}. 
We consider pressures of magnitude $P/k_\text{B} \sim 1 \  \text{K\AA}^{-2}$.
We also define the set $\boldsymbol{\alpha}=\{L_x,L_y,\theta\}$ of variational parameters characterizing the size and shape of the simulation cell. 
The trial wavefunction $\Psi_{\boldsymbol{\theta}}(\mathbf{R})$ depends in principle on all the variational parameters $\boldsymbol{\theta}$, with $\boldsymbol{\alpha}$ being a subset, that is $\boldsymbol{\alpha} \subset \boldsymbol{\theta}$.
However, in this section, we argue that it is much more convenient in practice to pass all the geometrical dependency $\boldsymbol{\alpha}$ to the Gibbs operator, and remove all of it from the trial wavefunction.
This can be achieved by sampling unit scaled coordinates $\mathbf{S} = (\mathbf{s}_1, \hdots, \mathbf{s}_N) \in ([0,1] \times [0,1])^N$ instead of the physical coordinates $\mathbf{R}=(\mathbf{r}_1,\hdots,\mathbf{r}_N)$. These physical coordinates can be obtained by the following affine transformation of the unit square
\begin{equation} \label{eq:affine_transformation_square_parallelogram}
    \begin{cases}
        x = L_x m + \cos \theta L_y n & \\
        y = \sin\theta L_y n &
    \end{cases}
    ,
\end{equation}
where $\mathbf{s} \equiv (m,n) \in [0,1]\times[0,1]$.
The affine transformation thus converts a (dimensionless unit) single-particle coordinate $\mathbf{s}_i$ to the associated physical single-particle coordinate $\mathbf{r}_i$, where $i$ is a particle index ranging from 1 to $N$.

\textbf{Gibbs energy estimator:}
In terms of the physical coordinates, the local Gibbs energy estimator $G_\text{loc}(\mathbf{R})$ is given by
\begin{align}
    G_\text{loc}(\mathbf{R}) 
    &= K_\text{loc}(\mathbf{R}) + V_\text{loc}(\mathbf{R}), \\
    K_\text{loc}(\mathbf{R})
    &= - \frac{1}{2} \sum_{i=1}^N \left[ \nabla_{\mathbf{r}_i}^2 \log \Psi(\mathbf{R}) + (\nabla_{\mathbf{r}_i} \log \Psi(\mathbf{R}))^2 \right], \label{eq:kinetic_estimator_physical_coordinates} \\
    V_\text{loc}(\mathbf{R}) 
    &= V(\mathbf{R}) + P L_x L_y \sin \theta. \label{eq:potential_estimator_physical_coordinates}
\end{align}
To derive an expression for the local energy estimator in terms of unit-scaled coordinates $\mathbf{S}$, it will be necessary to introduce generalized expressions for the Laplacian and the gradient to account for changes of coordinates.
In the language of differential geometry, a generalized expression for the Laplacian is provided by the Laplace-Beltrami operator 
\begin{equation} \label{eq:laplace_beltrami}
    \Delta = g^{\alpha \beta} \partial_\alpha \partial_\beta - g^{\alpha \beta} \Gamma_{\alpha \beta}^\gamma \partial_\gamma,
\end{equation}
where $g^{\alpha \beta}$ is the inverse metric tensor and $\Gamma_{\alpha \beta}^\gamma$ are the Christoffel symbols. The Einstein summation notation is used for the Greek indices. They range from 1 to $d$, with $d$ the spatial dimension. 
Here $d=2$ and we focus on the affine transformation described in \cref{eq:affine_transformation_square_parallelogram}.
The Jacobian $J_\alpha^\beta = \partial \mathbf{r}^\beta/\partial \mathbf{s}^\alpha$ of this transformation is
\begin{equation}
    J = 
    \begin{pmatrix}
        \frac{\partial x}{\partial m} & \frac{\partial x}{\partial n} \\
        \frac{\partial y}{\partial m} & \frac{\partial y}{\partial n}
    \end{pmatrix}
    =
    \begin{pmatrix}
        L_x & \cos \theta L_y \\
        0 & \sin \theta L_y
    \end{pmatrix}
    ,
\end{equation}
so that $\mathbf{r} = J \mathbf{s}$.
The Jacobian can then be used to define the metric, as can be seen by computing a dot product: $\mathbf{r} \cdot \mathbf{r} = \mathbf{r}^\text{T} \mathbf{r} = (J \mathbf{s})^\text{T} (J \mathbf{s}) = \mathbf{s}^\text{T} g \mathbf{s}$, where $g = J^T J$. 
The inverse metric is then given by $g^{-1} = J^{-1} (J^\text{T})^{-1} = J^{-1} (J^{-1})^\text{T}$. 
Alternatively, the metric can be defined with the basis set $\{\mathbf{e}_1, \mathbf{e}_2\}$, spanning a parallelogram, and its dual basis set $\{\mathbf{e}'^1, \mathbf{e}'^2\}$, given explicitly by
\begin{equation}
    \begin{cases}
        \mathbf{e}_1 = (L_x,0), & \\
        \mathbf{e}_2 = (\cos\theta L_y, \sin\theta L_y), &
    \end{cases}
    \qquad
    \begin{cases}
        \mathbf{e}'^1 = (1/L_x,-\text{cot}\theta/L_x), & \\
        \mathbf{e}'^2 = (0,\text{csc}\theta/L_y), &
    \end{cases}
\end{equation}
where the dual basis set is defined via $\mathbf{e}'^\alpha \mathbf{e}_\beta = \delta_\beta^\alpha$. 
The associated metric tensor $g_{\alpha \beta} = \mathbf{e}_\alpha \cdot \mathbf{e}_\beta$ and the inverse metric tensor $g^{\alpha \beta} = \mathbf{e}'^\alpha \cdot \mathbf{e}'^\beta$ for the dual space are respectively given by
\begin{equation}
    g_{\alpha \beta} = 
    \begin{pmatrix}
        L_x^2 & L_x L_y \cos\theta \\
        L_x L_y \cos \theta & L_y^2
    \end{pmatrix}
    ,
    \qquad
    g^{\alpha \beta} =
    \begin{pmatrix}
        \frac{ \text{csc}^2\theta }{ L_x^2} & -\frac{\text{cot}\theta \text{csc}\theta}{L_x L_y} \\
        -\frac{\text{cot}\theta \text{csc}\theta}{L_x L_y} & \frac{ \text{csc}^2\theta }{ L_y^2}
    \end{pmatrix}
    .
\end{equation}
In this particular case, the Christoffel symbols vanish (because the metric is independent of the coordinates) and the generalized Laplacian in \cref{eq:laplace_beltrami} reduces to $\Delta = g^{\alpha \beta} \partial_\alpha \partial_\beta$. 
We note that more ``exotic'' simulation cell geometries could be considered, introducing non-trivial Christoffel symbol contributions. 
In the particular case of a rectangular cell (with $\theta = \pi/2$), the inverse metric reduces to $g^{\alpha \beta} = \text{diag}(1/L_x^2, 1/L_y^2)$.

In turn, the gradient of the physical coordinates $\nabla_\mathbf{r} = (\partial_x, \partial_y)^\text{T}$ is obtained with the inverse Jacobian transposed, that is
\begin{equation}
    \nabla_\mathbf{r} = 
    \begin{pmatrix}
        \frac{\partial}{\partial x} \\
        \frac{\partial}{\partial y}
    \end{pmatrix}
    =
    \begin{pmatrix}
        \frac{\partial m}{\partial x} & \frac{\partial n}{\partial x} \\
        \frac{\partial m}{\partial y} & \frac{\partial n}{\partial y} \\
    \end{pmatrix}
    \begin{pmatrix}
        \frac{\partial}{\partial m} \\
        \frac{\partial}{\partial n}
    \end{pmatrix}
    = (J^{-1})^\text{T} \nabla_\mathbf{s}
\end{equation}
and the inverse Jacobian reads
\begin{equation}
    J^{-1} = 
    \begin{pmatrix}
        \frac{\partial m}{\partial x} & \frac{\partial m}{\partial y} \\
        \frac{\partial n}{\partial x} & \frac{\partial n}{\partial y} \\
    \end{pmatrix}
    =
    \begin{pmatrix}
        \frac{1}{L_x} & - \frac{\cot \theta}{L_x} \\
        0 & \frac{\csc \theta}{L_y}
    \end{pmatrix}
    .
\end{equation}

\begin{figure}
    \centering
    \includegraphics[width=0.5\textwidth]{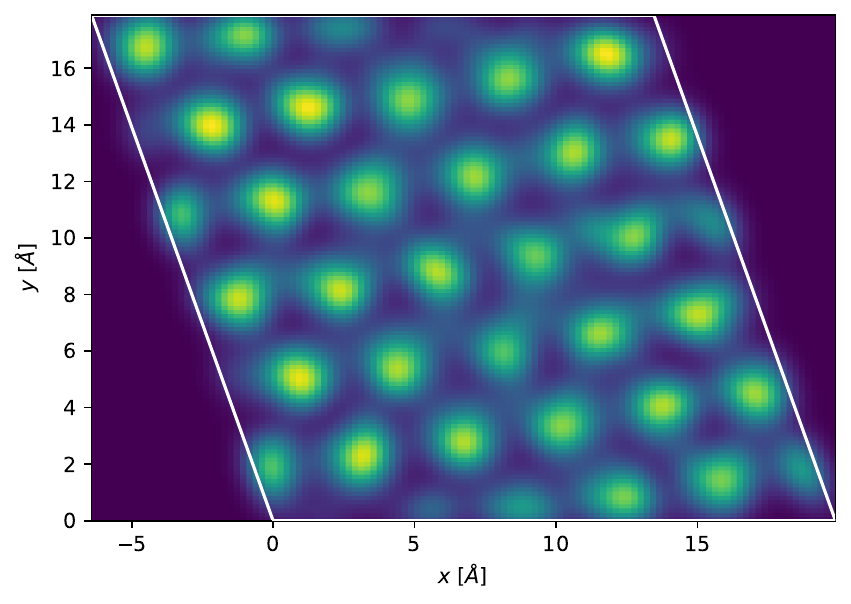}
    \caption{The density operator of the ground state of $N=31$ helium-4 atoms is shown in the isobaric ensemble at a fixed pressure of $P = 2.4$ K\AA$^{-2}$.
    The simulation box geometry and shape, characterised by $L_x$, $L_y$ and $\theta$, are then treated as variational parameters.
    By optimizing these parameters, the minimal energy configuration can be found even with an odd $N$. 
    The white lines mark the boundary of the simulation cell.
    The Monte Carlo samples used to calculate the density operator come from the same Markov chain (else no signal would be found because the trial wavefunction is translation invariant). 
    }
    \label{fig:density_operator_he4_N=31_d=2_p=2.4}
\end{figure}

As a result, the kinetic and potential energy parts of the local Gibbs energy estimator can be written in terms of the unit coordinates as follows 
\begin{align} 
    \tilde{K}_\text{loc}(\mathbf{S}) 
    &= -\frac{1}{2} \sum_{i=1}^N \left[ \sum_{\alpha,\beta=1}^d g^{\alpha \beta} \frac{ \partial^2 }{\partial \mathbf{s}_i^\alpha \partial \mathbf{s}_i^\beta} \log \Psi(\mathbf{S}) + ((J^{-1})^\text{T} \nabla_{\mathbf{s}_i} \log \Psi(\mathbf{S}) )^2 \right], \\
    \tilde{V}_\text{loc}(\mathbf{S})
    &= V(J \mathbf{S}) + P L_x L_y \sin \theta,
\end{align}
where $J \mathbf{S} \equiv (J \mathbf{s}_1, \hdots, J \mathbf{s}_N)$.
While these expressions look more complicated than \cref{eq:kinetic_estimator_physical_coordinates,eq:potential_estimator_physical_coordinates}, they simplify the computer implementation since the Monte Carlo configurations $\mathbf{S}$ can be sampled from the unit box $([0,1] \times [0,1])^N$ independently of the varying geometrical parameters $L_x$, $L_y$ and $\theta$.
We note that, in practice, the following dimensionless Gibbs energy estimator is implemented
\begin{equation} \label{eq:gibbs_energy_scaled_coordinates}
    \tilde{G}_\text{loc}(\mathbf{S}) = \frac{\hbar^2}{m r_m^2} \left[ r_m^2 \tilde{K}_\text{loc}(\mathbf{S}) + \frac{\epsilon m r_m^2}{\hbar^2} \left( V( J \mathbf{S}) + \frac{P}{\epsilon} L_x L_y \sin \theta \right) \right],
\end{equation}
where $\epsilon/k_B = 10.8 \ \text{K}$ is the well depth parameter extracted from the Aziz potential (see \cref{appendix:potential_and_pbcs}).
In case where the chosen pressure would result in the formation of a triangular lattice, the angle $\theta$ of the simulation cell can be fixed to $\pi/2$ if $N$ is even, because  the simulation cell can be tiled with an integer multiple of the unit cell.
However, if $N$ is odd, which disrupts the periodicity, $\theta$ can be left as a free parameter to reach a simulation cell geometry that would be commensurate with the triangular lattice unit cell.
The density operator of such a state is shown in \cref{fig:density_operator_he4_N=31_d=2_p=2.4}. 
While the crystal structure is simple for 2D helium-4, helium isotopes in 3D can adopt different crystal configurations based on the pressure applied on the sample.
The method presented here would facilitate the study of such systems, allowing to learn the minimal energy configuration instead of imposing it.

\textbf{Updating the variational parameters:}
First, we define the set of all variational parameters excluding $\boldsymbol{\alpha}=\{L_x,L_y,\theta\}$ as $\mathbf{W}=\boldsymbol{\theta} \setminus \boldsymbol{\alpha}$.
By sampling unit scaled configurations $\mathbf{S}$ instead of the physical ones $\mathbf{R}$, we use the following trial wavefunction to compute expectation values:
\begin{equation}
    \Phi_\mathbf{W}(\mathbf{S}) \equiv \Psi_{\mathbf{W} \cup \{L_x=1,L_y=1,\theta=\pi/2\}}(J^{-1}\mathbf{R}),
\end{equation}
instead of the one defined in physical space, i.e. $\Psi_{\boldsymbol{\theta}}(J \mathbf{S}) = \Psi_{\boldsymbol{\theta}}(\mathbf{R})$ (see \cref{appendix:helium_ground_state_implementation} for its implementation).
We further assumed that $N$ was even to fix $\theta$ to $\pi/2$.
As a result, the gradient of the Gibbs operator $\tilde{G}(\boldsymbol{\alpha})$ with respect to the geometrical variational parameters is 
\begin{equation}
    \nabla_{\boldsymbol{\alpha}} \langle \tilde{G}(\boldsymbol{\alpha}) \rangle = \nabla_{\boldsymbol{\alpha}} \frac{\langle \Phi_\mathbf{W} | \tilde{G}(\boldsymbol{\alpha}) | \Phi_\mathbf{W} \rangle}{\langle \Phi_\mathbf{W}| \Phi_\mathbf{W} \rangle} = \mathbb{E}[\nabla_{\boldsymbol{\alpha}} \tilde{G}_\text{loc}(\mathbf{S})],
\end{equation}
where the statistical average is over the Born distribution $|\Phi_\mathbf{W}(\mathbf{S})|^2$, and, as in \cref{eq:gibbs_energy_scaled_coordinates}, we dropped in the last step the explicit dependence of $\tilde{G}_\text{loc}(\mathbf{S})$ on $\boldsymbol{\alpha}$. The quantity $\nabla_{\boldsymbol{\alpha}} \tilde{G}_\text{loc}(\mathbf{S})$ was calculated with automatic differentiation, and, importantly, the force term in \cref{eq:gradient_wrt_a_var_param} is not present (because by construction $\Phi_\mathbf{W}(\mathbf{S})$ is independent of $\boldsymbol{\alpha}$).
On the other hand, the gradient of $\langle \tilde{G}(\boldsymbol{\alpha}) \rangle$ with respect to the rest of the variational parameters $\mathbf{W}$ only involves this force term.

\textbf{Stochastic reconfiguration:}
We implicitly defined above $\boldsymbol{\theta} \equiv (\mathbf{W},\boldsymbol{\alpha})$ as the vector containing all the variational parameters. 
The update $\boldsymbol{\delta \theta}$ of these parameters is found by solving the following set of equations
\begin{equation} \label{eq:sr_set_of_equations}
    \sum_{k'} \Tilde{S}_{k k'} \boldsymbol{\delta \theta}_{k'} = - \boldsymbol{\eta}_k \mathbf{F}_k,
\end{equation}
where $\Tilde{S}_{k k'} = S_{k k'} + \delta_{k k'} \boldsymbol{\epsilon}_k$, with the $S$-matrix entries given by
\begin{equation}
    S_{k k'} = \frac{ \langle \partial_{\boldsymbol{\theta}_k} \Phi_{\mathbf{W}} | \partial_{\boldsymbol{\theta}_{k'}} \Phi_{\mathbf{W}} \rangle }{\langle \Phi_{\mathbf{W}} | \Phi_{\mathbf{W}} \rangle } 
    -
    \frac{ \langle \partial_{\boldsymbol{\theta}_k} \Phi_{\mathbf{W}} | \Phi_{\mathbf{W}} \rangle \langle \Phi_{\mathbf{W}} | \partial_{\boldsymbol{\theta}_{k'}} \Phi_{\mathbf{W}} \rangle }{ \langle \Phi_{\mathbf{W}} | \Phi_{\mathbf{W}} \rangle^2 }.
\end{equation}

The vector $\boldsymbol{\epsilon}_k$ is a diagonal shift introduced for numerical stability, $\boldsymbol{\eta}_k$ is the learning rate and $\mathbf{F}_k$ is the gradient of the loss function (discussed in \cref{appendix:gradient_estimator}).
We choose the diagonal shift and the learning rate vectors to have the following form
\begin{align}
    \boldsymbol{\epsilon} &\equiv (1,1,1,\epsilon,\hdots,\epsilon), \\
    \boldsymbol{\eta} &\equiv (\Delta L_x, \Delta L_y, \Delta \theta, \eta, \hdots, \eta),
\end{align}
so that the diagonal shift and the learning rate are fixed to a constant for all variational parameters, respectively $\epsilon$ and $\eta$, except for the entries corresponding to $L_x$, $L_y$ and $\theta$.
In the $\boldsymbol{\alpha}$ subspace, the $S$-matrix vanishes, i.e.
\begin{equation}
    S|_{\boldsymbol{\alpha}} = 
    \hspace{1mm}
    \begin{blockarray}{cccc}
    L_x & L_y & \theta \\
    \begin{block}{(ccc)c}
    0 & 0 & 0 & \hspace{1mm} L_x \\
    0 & 0 & 0 & \hspace{1mm} L_y \\
    0 & 0 & 0 & \hspace{1mm} \theta \\
    \end{block}
    \end{blockarray}
    .
\end{equation}
Hence \cref{eq:sr_set_of_equations} reduces to standard stochastic gradient descent for the parameters $L_x$, $L_y$ and $\theta$. 
However, more advanced optimizers like Adam can still be used for these three parameters.

\section{R\'enyi-2 entropy}

\subsection{Standard swap estimator}

Given two partitions A and B of a physical system, the reduced density matrix of subsystem A is obtained via $\rho_\text{A} = \text{Tr}_\text{B} \rho$, that is, by tracing out the degrees of freedom of subsystem B.
Here we will focus on the case where the full system is in a pure quantum state $| \Psi \rangle$, so that the density matrix is $\rho = | \Psi \rangle \langle \Psi | / \langle \Psi | \Psi \rangle$. 
To simplify notation we will assume that $|\Psi\rangle$ has unit norm (i.e. $\langle \Psi | \Psi \rangle = 1$).
One can then measure the bipartite entanglement using the R\'enyi-2 entropy
\begin{equation}
    S_2[\rho_\text{A}] = - \ln(\text{Tr}_\text{A} \rho_\text{A}^2).
\end{equation}
Monte Carlo estimators for $\text{Tr}_\text{A} \rho_\text{A}^2$ were presented for instance in \cite{2010_hastings_renyi2,2012_tubman_renyi2_vmc, 2014_herdman_renyi2_pimc,2023_sinibaldi_unbiasing,2024_mauron_topological_entanglement}.
We will quickly review how to proceed for a system of $N$ particles in $d$-dimensional continuous space.
To do so, we first split the matrix of single-particle coordinates $\mathbf{R} \in \mathbb{R}^{N \times d}$ in two subsets, that is $\mathbf{R} = \mathbf{R}_\text{A} \cup \mathbf{R}_\text{B}$, where $\mathbf{R}_\text{A}$ ($\mathbf{R}_\text{B}$) is the set of single-particle coordinates in partition A (B).
By tracing out subsystem B, we get 
\begin{align}
    \langle \mathbf{R}_\text{A} | \rho_\text{A} | \mathbf{R}_\text{A}'' \rangle 
    &= \int d \mathbf{R}_\text{B} \langle \mathbf{R}_\text{A} \mathbf{R}_\text{B} | \rho | \mathbf{R}_\text{A}'' \mathbf{R}_\text{B} \rangle  \\
    &= \int d \mathbf{R}_\text{B} \Psi^*(\mathbf{R}_\text{A},\mathbf{R}_\text{B}) \Psi(\mathbf{R}_\text{A}'',\mathbf{R}_\text{B}),
\end{align}
so that the matrix elements of $\rho_\text{A}^2$ are given by
\begin{align}
    \langle \mathbf{R}_\text{A} | \rho_\text{A}^2 | \mathbf{R}_\text{A}'' \rangle
    &= \int d \mathbf{R}_\text{A}' \langle \mathbf{R}_\text{A} | \rho_\text{A} | \mathbf{R}_\text{A}' \rangle \langle \mathbf{R}_\text{A}' | \rho_\text{A} | \mathbf{R}_\text{A}'' \rangle \\
    &= \int d \mathbf{R}_\text{A}' d \mathbf{R}_\text{B} d \mathbf{R}_\text{B}' \Psi^*(\mathbf{R}_\text{A}, \mathbf{R}_\text{B}) \Psi(\mathbf{R}_\text{A}',\mathbf{R}_\text{B}) \Psi^*(\mathbf{R}_\text{A}',\mathbf{R}_\text{B}') \Psi(\mathbf{R}_\text{A}'',\mathbf{R}_\text{B}').
\end{align}
The domains of integration (either partition A or partition B) are left implicit but they can be deduced from the name of the dummy variables. We can then trace over subsystem A to get
\begin{align} \label{eq:tr_rho_A_square}
    \text{Tr}_\text{A} \rho_\text{A}^2 
    &= \int d \mathbf{R}_\text{A} \langle \mathbf{R}_\text{A} | \rho_\text{A}^2 | \mathbf{R}_\text{A} \rangle \\
    &= \int d \mathbf{R}_\text{A} d \mathbf{R}_\text{A}' d \mathbf{R}_\text{B} d \mathbf{R}_\text{B}' \Psi^*(\mathbf{R}_\text{A}, \mathbf{R}_\text{B}) \Psi(\mathbf{R}_\text{A}',\mathbf{R}_\text{B}) \Psi^*(\mathbf{R}_\text{A}',\mathbf{R}_\text{B}') \Psi(\mathbf{R}_\text{A},\mathbf{R}_\text{B}').
\end{align}
The so-called swap estimator is then defined in the following way
\begin{align} \label{eq:standard_estimator}
    \text{Tr}_\text{A} \rho_\text{A}^2 
    &= \int d\mathbf{R}_\text{A} d\mathbf{R}_\text{A}' d\mathbf{R}_\text{B} d\mathbf{R}_\text{B}' | \Psi(\mathbf{R}_\text{A}, \mathbf{R}_\text{B}) |^2 | \Psi(\mathbf{R}_\text{A}',\mathbf{R}_\text{B}') |^2 \left[ \frac{ \Psi(\mathbf{R}_\text{A}', \mathbf{R}_\text{B}) \Psi(\mathbf{R}_\text{A}, \mathbf{R}_\text{B}') }{ \Psi(\mathbf{R}_\text{A}, \mathbf{R}_\text{B}) \Psi(\mathbf{R}_\text{A}', \mathbf{R}_\text{B}')} \right] \nonumber\\
    &= \mathbb{E} \left[ \frac{ \Psi(\mathbf{R}_\text{A}', \mathbf{R}_\text{B}) \Psi(\mathbf{R}_\text{A}, \mathbf{R}_\text{B}') }{ \Psi(\mathbf{R}_\text{A}, \mathbf{R}_\text{B}) \Psi(\mathbf{R}_\text{A}', \mathbf{R}_\text{B}')} \right],
\end{align}
Two configurations $\mathbf{R}$ and $\mathbf{R}'$, sampled from the ground state Born distribution, are thus necessary to evaluate this estimator. 
If the number of particles in partition A (and hence in partition B) is not the same for the two configurations, the estimator vanishes, i.e. $\text{Tr}_\text{A} \rho_\text{A}^2=0$, since the numerator in \cref{eq:standard_estimator} evaluates to 0.

\begin{figure}[!h]
    \centering
    \includegraphics[width=0.7\textwidth]{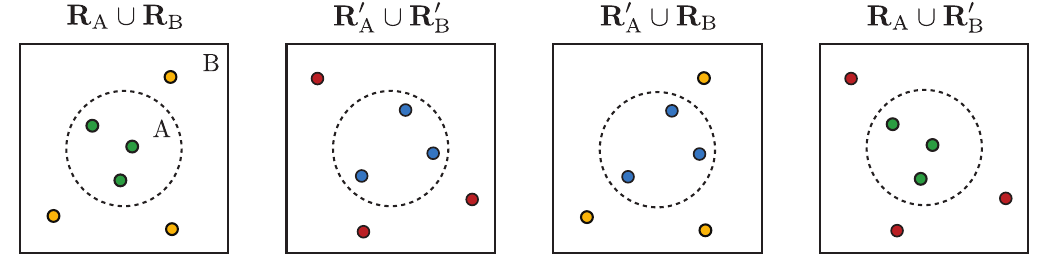}
    \caption{
    Sketch of the four particle configurations necessary to evaluate the estimator of the R\'enyi-2 entropy given in \cref{eq:standard_estimator}.
    Partition A is defined by a circle of radius $R$ centered in the middle of the simulation cell (with its boundary shown as a dashed line). 
    Partition B is the complement of A.
    The configurations $\mathbf{R}_\text{A} \cup \mathbf{R}_\text{B}$ and $\mathbf{R}_\text{A}' \cup \mathbf{R}_\text{B}'$ both have the same total number of particles ($N=6$) and the same number of particles in partition A ($N_\text{A}=3$). 
    It is therefore possible to swap their coordinates, and so $\text{Tr}_\text{A} \rho_\text{A}^2 \neq 0$.
    }
    \label{fig:tr_rho_A2_estimator_sketch}
\end{figure}

\subsection{Truncated swap estimator}

To simplify notation, we let $\mathbf{x}_1 \equiv \mathbf{R}_\text{A} \cup \mathbf{R}_\text{B}, \mathbf{x}_2 \equiv \mathbf{R}_\text{A}' \cup \mathbf{R}_\text{B}', \mathbf{y}_1 \equiv \mathbf{R}_\text{A}' \cup \mathbf{R}_\text{B}$
 and $\mathbf{y}_2 \equiv \mathbf{R}_\text{A} \cup \mathbf{R}_\text{B}'$.
We also denote the metric of integration by $d \Omega \equiv d\mathbf{R}_\text{A} d\mathbf{R}_\text{A}' d\mathbf{R}_\text{B} d\mathbf{R}_\text{B}'$. \cref{eq:tr_rho_A_square} can then be written more compactly as follows
\begin{equation}
    \text{Tr}_\text{A} \rho_\text{A}^2 = \int d \Omega \Psi^*(\mathbf{x}_1) \Psi^*(\mathbf{x}_2) \Psi(\mathbf{y}_1) \Psi(\mathbf{y}_2).
\end{equation}

Outliers can often skew the distribution of swap estimators.
To eliminate their effect, we use an idea from \cite{2010_van_den_nest_simulating_quantum_computers_probabilistic} to truncate the swap estimator values with the following step function
\begin{equation}
    \varepsilon(\mathbf{x},\mathbf{y}) \equiv \varepsilon(\{\mathbf{x}_1,\mathbf{x}_2\}, \{\mathbf{y}_1,\mathbf{y}_2\}) 
    = \begin{cases}
        1, & \left| \frac{ \Psi(\mathbf{y}_1) \Psi(\mathbf{y}_2) }{ \Psi(\mathbf{x}_1) \Psi(\mathbf{x}_2) } \right| < 1, \\
        0, & \text{else},
    \end{cases}
\end{equation}
and noticing that the following identity holds
\begin{equation} \label{eq:truncation_identity}
    \varepsilon(\mathbf{x},\mathbf{y}) + \varepsilon(\mathbf{y},\mathbf{x}) + \delta_{\left| \frac{ \Psi(\mathbf{y}_1) \Psi(\mathbf{y}_2) }{ \Psi(\mathbf{x}_1) \Psi(\mathbf{x}_2) } \right|, 1} = 1.
\end{equation}
Indeed, this last identity takes into account the three possible outcomes of the SWAP ratio: either it is (1) less than 1, (2) greater than 1 or (3) equal to 1. 
By using \cref{eq:truncation_identity}, the quantity $\text{Tr}_\text{A} \rho_\text{A}^2$ can be exactly rewritten as a sum of two statistical averages
\begin{align}
    \text{Tr}_\text{A} \rho_\text{A}^2 &
    = \int d \Omega \Psi^*(\mathbf{x}_1) \Psi^*(\mathbf{x}_2) \Psi(\mathbf{y}_1) \Psi(\mathbf{y}_2) \left[ \varepsilon(\mathbf{x},\mathbf{y}) + \varepsilon(\mathbf{y},\mathbf{x}) + \delta_{\left| \frac{ \Psi(\mathbf{y}_1) \Psi(\mathbf{y}_2) }{ \Psi(\mathbf{x}_1) \Psi(\mathbf{x}_2) } \right|, 1} \right] \nonumber\\
    &= \int d \Omega |\Psi(\mathbf{x}_1)|^2 |\Psi(\mathbf{x}_2)|^2 \left[ \frac{ \Psi(\mathbf{y}_1) \Psi(\mathbf{y}_2) }{ \Psi(\mathbf{x}_1) \Psi(\mathbf{x}_2) } \right] \varepsilon(\mathbf{x},\mathbf{y}) \nonumber\\
    & \quad + \int d \Omega |\Psi(\mathbf{y}_1)|^2 |\Psi(\mathbf{y}_2)|^2 \left[ \frac{ \Psi^*(\mathbf{x}_1) \Psi^*(\mathbf{x}_2) }{ \Psi^*(\mathbf{y}_1) \Psi^*(\mathbf{y}_2) } \right] \varepsilon(\mathbf{y},\mathbf{x}) \nonumber\\
    & \quad + \int d \Omega |\Psi(\mathbf{x}_1)|^2 |\Psi(\mathbf{x}_2)|^2 \left[ \frac{ \Psi(\mathbf{y}_1) \Psi(\mathbf{y}_2) }{ \Psi(\mathbf{x}_1) \Psi(\mathbf{x}_2) } \right]  \delta_{\left| \frac{ \Psi(\mathbf{y}_1) \Psi(\mathbf{y}_2) }{ \Psi(\mathbf{x}_1) \Psi(\mathbf{x}_2) } \right|, 1} \nonumber \\
    &= \int d \Omega |\Psi(\mathbf{x}_1)|^2 |\Psi(\mathbf{x}_2)|^2 \left[ \frac{ \Psi(\mathbf{y}_1) \Psi(\mathbf{y}_2) }{ \Psi(\mathbf{x}_1) \Psi(\mathbf{x}_2) } \right] \varepsilon(\mathbf{x},\mathbf{y}) \nonumber\\
    & \quad + \int d \Omega |\Psi(\mathbf{x}_1)|^2 |\Psi(\mathbf{x}_2)|^2 \left[ \frac{ \Psi^*(\mathbf{y}_1) \Psi^*(\mathbf{y}_2) }{ \Psi^*(\mathbf{x}_1) \Psi^*(\mathbf{x}_2) } \right] \varepsilon(\mathbf{x},\mathbf{y}) \nonumber\\
    & \quad + \int d \Omega |\Psi(\mathbf{x}_1)|^2 |\Psi(\mathbf{x}_2)|^2 \left[ \frac{ \Psi(\mathbf{y}_1) \Psi(\mathbf{y}_2) }{ \Psi(\mathbf{x}_1) \Psi(\mathbf{x}_2) } \right]  \delta_{\left| \frac{ \Psi(\mathbf{y}_1) \Psi(\mathbf{y}_2) }{ \Psi(\mathbf{x}_1) \Psi(\mathbf{x}_2) } \right|, 1} \nonumber \\
    &= \mathbb{E} \left[ 2 \text{Re} \left[ \frac{ \Psi(\mathbf{y}_1) \Psi(\mathbf{y}_2) }{ \Psi(\mathbf{x}_1) \Psi(\mathbf{x}_2) } \right] \epsilon(\mathbf{x},\mathbf{y}) \right]
    + \mathbb{E} \left[ e^{i \text{Arg}[ \Psi(\mathbf{y}_1) \Psi(\mathbf{y}_2) \Psi^*(\mathbf{x}_1) \Psi^*(\mathbf{x}_2)]} \delta_{\left| \frac{ \Psi(\mathbf{y}_1) \Psi(\mathbf{y}_2) }{ \Psi(\mathbf{x}_1) \Psi(\mathbf{x}_2) } \right|, 1}  \right],
\end{align}
where the probability distribution for both statistical averages is a product of two independent ground state Born distributions.
Note that we relabeled $\mathbf{x}$ and $\mathbf{y}$ in the second term of  and grouped the first two terms (using $z_1^* z_2^* = (z_1 z_2)^*$ and then $z+z^*=2 \text{Re}[z]$, for any $z_1,z_2,z \in \mathbb{C}$). In the third term, the magnitude of the ratio is 1 so only the phase survives (and we then used that $\text{Arg}[z_1]+\text{Arg}[z_2] = \text{Arg}[z_1 z_2]$ for any $z_1,z_2 \in \mathbb{C}$).

\subsection{Periodic systems and random origins}

Periodic systems do not have a preferred origin.
As a result, for each Monte Carlo sample, we randomly displace the center of partition A in the simulation cell, denoted $\mathcal{C} \equiv [0,L_x] \times [0,L_y]$, to calculate different instances of the $\text{Tr}_\text{A} \rho_\text{A}^2$ estimator.
For 2D helium-4, this averaging over different origins is especially relevant in the solid phase, where the triangular lattice forms.
In particular, let $\mathbf{o} \sim \mathcal{U}(\mathcal{C})$ be the origin associated to a given Monte Carlo sample $\mathbf{x}_1 = \mathbf{R}_\text{A} \cup \mathbf{R}_\text{B}$, where $\mathcal{U}(\mathcal{C})$ denotes a uniform distribution over the simulation cell $\mathcal{C}$. 
The circle of radius $R$ and center $\mathbf{o}$ then defines partition A.
The set of single-particle coordinates in partition A (and in the Monte Carlo sample $\mathbf{x}_1$) is thus given by 
\begin{equation}
\mathbf{R}_\text{A} \equiv \left\{\mathbf{r}_i \in \mathbf{x}_1 : d_\mathrm{mic}(\mathbf{r}_i, \mathbf{o}) < R \right\},
\end{equation}
where the minimum image convention distance $d_\mathrm{mic}$ was introduced in \cref{eq:minimum_image_convention_distance}.
Partition B is defined as the complement of partition A, i.e. $\mathbf{R}_\text{B} = \mathbf{R} \setminus \mathbf{R}_\text{A}$.
Another origin $\mathbf{o}'$ can be randomly selected for a second sample $\mathbf{x}_2 = \mathbf{R}_\text{A}' \cup \mathbf{R}_\text{B}'$.
The swapped configurations are then given by 
\begin{align}
    \mathbf{y}_1 &= (\mathbf{R}_\text{A}' - \mathbf{o}' + \mathbf{o}, \mathbf{R}_\text{B}), \\
    \mathbf{y}_2 &= (\mathbf{R}_\text{A} - \mathbf{o} + \mathbf{o}', \mathbf{R}_\text{B}').
\end{align}

\end{document}